\documentclass[11pt]{article}
\usepackage{amsmath,amssymb}
\usepackage{graphics}
\usepackage{graphicx}
\newcommand{\spc}{\hspace{0.22in}}

\newcommand{\vts}{\hspace{0.07cm}}
\newif\ifequationnumber
\def\eqswitchon{\equationnumbertrue
\@addtoreset{equation}{section}
\def\theequation{\arabic{section}.\arabic{equation}}
}
\def \ds {\displaystyle}

\def \ns {\normalsize}
\def \sp {\scriptsize}
\def \es {\enspace}
\def \ts {\thinspace}
\def \nts {\negthinspace}

\title{
  \vspace*{-0.5cm}
  \hfill{\ns KEK-TH-787} \\ 
  \vspace*{-0.5cm}
  \hfill{\ns hep-lat/0304011} \\ 
  \vspace*{-0.5cm}
  \hfill{\ns April 2003}  \\
  \vspace*{2.0cm}
  {\LARGE {\bf Dynamical Regge Calculus}} \\
  \vspace*{0.3cm}
  {\LARGE {\bf as Lattice Quantum Gravity}} \\
  \vspace*{0.8cm}
}
\author{ {\sc Hiroyuki Hagura}\thanks{e-mail: hiroyuki.hagura@kek.jp} }
\date{}
\begin{document}
\maketitle
\begin{center}
{\it Radiation Science Center, \\
High Energy Accelerator Research Organization (KEK), \\
Oho 1-1 Tsukuba-shi Ibaraki 305-0801, Japan}
\end{center}
\vspace{1cm} 
%
%
\begin{abstract}
  We propose a hybrid model of simplicial quantum gravity by performing at once
dynamical triangulations and Regge calculus. A motive for the hybridization is
to give a dynamical description of topology-changing processes of Euclidean 
spacetime. In addition, lattice diffeomorphisms as invariance of the simplicial
geometry are generated by certain elementary moves in the model. We attempt 
also a lattice-theoretic derivation of the black hole entropy 
using the symmetry.
Furthermore, numerical simulations of 3D pure gravity are carried out, 
exhibiting a large hysteresis between two phases.
We also measure geometric properties of Euclidean `time slice' based on a 
geodesic distance, resulting in a fractal structure in the 
strong-coupling phase. Our hybrid model not only reproduces numerical results 
consistent with those of dynamical triangulations and Regge calculus, but also 
opens a possibility of studying quantum black hole physics on the lattice.
\end{abstract}
\newpage
%
\section{Introduction}\label{sect:introduction}
\spc One of the most difficult problems in modern physics is the construction 
of a consistent theory of quantum gravity, whereas Einstein's theory of 
classical gravity has been very successful in explaining the large scale 
structure of spacetime~\cite{HawkingEllis}. Many programs to formulate the 
full quantum theory of gravity are under active 
research~\cite{stringText,Rovelli,David}.
Since none of them are decisive at present, it is still an open problem to 
understand what is quantum spacetime at microscopic scales.

  In this paper we explore a possibility that dynamical Regge 
calculus~\cite{DRC,DRC2}, which is a hybrid lattice model of dynamical 
triangulations~\cite{Weingarten,DT} and quantum Regge calculus~\cite{HamRev}, 
gives a possible candidate for a constructive definition of quantum gravity. 
Although traditional approaches, namely, Regge calculus and dynamical 
triangulations, have been well studied for a long time as lattice field 
theories of gravity, they are not satisfactory in several respects.
A necessary enlargement of physical degrees of freedom in lattice gravity 
strongly motivates our study on the hybridization. Actually, the enlargement 
enables us to define exact ``diffeomorphism-invariance'' on the lattice at 
least classically~\cite{DRC}. 

  We will apply the symmetry to a lattice-theoretic derivation~\cite{DRC2} of 
the Bekenstein-Hawking entropy of a black hole. 
By the end of 1970's it was 
generally accepted that a black hole has the huge entropy ${\cal S}_{\rm BH}$ 
proportional to its horizon area $A$~\cite{Bekenstein,Hawking}:
\begin{eqnarray}
{\cal S}_{\rm BH} = \frac{\ds 1}{\ds 4} k_{\rm B} 
\frac{\ds c^3 \! A}{\ds \hbar \hspace{0.5mm} G} \es,
\label{eq:entropyBH}
\end{eqnarray}
where $k_{\rm B}$ is Boltzmann's constant, $c$ the speed of light and $G$ 
Newton's constant\footnote{We write down explicitly all the constants in 
eq.~(\ref{eq:entropyBH}) in order to show how large the black hole entropy is. 
Indeed, for a black hole of horizon area $A = 1 \,{\rm cm^2}$, one obtains 
a huge value ${\cal S}_{\rm BH}/k_{\rm B} \sim 10^{65}$.}. 
Nowadays, it is generally believed that if one sticks to usual theories of
gravity (e.g.~Einstein's gravity and dilaton type gravity coupled to normal 
matter fields), eq.~(\ref{eq:entropyBH}) is widely valid~\cite{MBKOP}. 
Theoretical explanation of the origin of the huge entropy 
(\ref{eq:entropyBH}) is considered to be a necessary ingredient for any 
consistent theory of quantum gravity~\cite{Bekenstein2}.

  Another motive for studying the hybrid model is giving a description of
topology-changing processes of Euclidean spacetime in a dynamical 
way. In a continuum approach Hawking calculated semiclassically the 
contribution of gravitational instantons to such processes of 
four-dimensional Euclidean spacetime manifolds~\cite{foamHawking} and
discussed the phenomenon in connection with Regge calculus; the resultant 
spacetime that is highly curved and has all possible topologies is called the 
{\em spacetime foam}. Such a complicated spacetime is expected to appear in the
strong-coupling region of quantum gravity~\cite{spcTmFoam}, especially inside 
black holes and in the very early universe. Thus, a question naturally 
arises: How can we describe the spacetime foam in lattice quantum gravity?
It will be shown that the topology-changing processes can in principle 
occur via degenerate simplicial configurations in our hybrid model.
  
  Regge's lattice formulation of general relativity~\cite{Regge} has been 
applied to quantum gravity in two different manners, that is, quantum Regge 
calculus and dynamical triangulations, as mentioned above.
In the former, all link-lengths in a fixed pattern of triangulation (a fixed
connectivity of vertices) play the role of dynamical variables instead of
the metric field $g_{\mu\nu}$. Accordingly, the integration of the link-lengths
with a proper functional measure is assumed to give a constructive definition 
of the path integral for the field $g_{\mu\nu}$. In the latter, in contrast, 
all patterns of triangulations are regarded as dynamical, while all the 
link-lengths are fixed to a single lattice spacing $a$. In this case, the sum 
over all possible triangulations is expected to give another constructive 
definition of the path integral for the gravitational field.

  Both lattice models of gravity have merits and demerits; the weakest point 
common to both of them is the lack of the gauge symmetry that corresponds to 
`reparametrization-invariance' on the lattice.
Actually, one must first give a precise definition of the 
`reparametrization-invariance' on the lattice in order to construct a 
`gauge-invariant' measure. We call the symmetry {\em lattice 
diffeomorphism-invariance}. Physically, the invariance should be the lattice 
counterpart of the principle of general covariance.
%
   An intention of this paper is to discuss the lattice symmetry even on 
the finite lattice by covering as large degrees of freedom as possible
in the space of Riemannian geometries. Thus, one would conjecture that such an 
enlargement of degrees should be realized by performing {\em at once\/} the 
link-integration arising from quantum Regge calculus and the triangulation-sum 
arising from dynamical triangulations. 

   The plan of this paper is as follows. In section~\ref{sect:reviews} we 
give a short review of quantum Regge calculus and dynamical triangulations, 
making a brief comparison between them. How each formulation regularizes the 
space of Riemannian geometries is the central issue there. 
Section~\ref{sect:defDRC} deals with the definition of the hybrid model, 
that is, dynamical Regge calculus, and discuss its fundamental properties. 
It will also be discussed that the topology change of discrete spacetime can 
occur in principle via degenerate simplicial complexes, because the lattice 
action for the gravitational field remains well-defined and finite on any 
degenerate configuration that is {\em not\/} a manifold.
  Section~\ref{sect:genCovLat} considers the symmetric property, that is,
the lattice diffeomorphism-invariance. We will also attempt a
lattice-theoretic derivation of the black hole entropy (\ref{eq:entropyBH}) 
using the lattice symmetry and give a simple interpretation of it in connection
with the spacetime uncertainty principle of string theory~\cite{yoneya,NCG,
GKP}. Section~\ref{sect:numSimln} is devoted to numerical studies of 
three-dimensional pure gravity. In particular, a fractal structure based on a 
geodesic distance~\cite{KKMW} is measured in detail. As a result, we will 
obtain a picture of quantum spacetime, which is similar to that
obtained in DT. Furthermore, we will acquire several other numerical results 
consistent with those of both dynamical triangulations and 
quantum Regge calculus. In section~\ref{sect:sumryDscsn} we will summarize 
our results and discuss several perspectives of dynamical Regge calculus.
\vspace*{0.5cm}
%
\section{Quantum Regge calculus vs. dynamical triangulations}
\label{sect:reviews}
\spc In this section we will make a comparison between the two formulations of 
lattice gravity. We first explain briefly the idea of the Euclidean path 
integral approach in the continuum, and then present a basics of Regge 
calculus. Emphasis is laid on geometric properties on the finite lattice.
%
\subsection{Regge's formulation}\label{sect:ReggeFormulation}
\spc Leaving aside the lattice approaches for a moment, let us look at 
the Euclidean path-integral approach in the continuum~\cite{HawkingIs}.
The basic object of the approach is the functional integral for the metric 
field $g_{\mu\nu}$ of Euclidean signature $+ + \cdots +$:
\begin{eqnarray}
Z_{\rm cont} = \int 
\frac{\ds {\cal D} g}{\ds \mbox{Vol} (\mbox{Diff}_M)} \ts
\exp \Bigl( -S_{\rm EH}[g] \Bigr) \es, 
\label{eq:Z_cont}
\end{eqnarray}
where $S_{\mbox{\sp EH}}[g]$ is the Einstein-Hilbert action on a 
$d$-dimensional Euclidean spacetime $M$:
\begin{eqnarray}
S_{\rm EH}[g] = - \frac{\ds 1}{\ds 16\pi G}
\int_{M} \mbox{d}^d x \sqrt{g} \Bigl( R - 2{\mit\Lambda} \Bigr) \es.
\label{eq:S_EH}
\end{eqnarray}
Here ${\mit\Lambda}$ is the cosmological constant, and units are such that 
$c = \hbar = 1$. For simplicity, we assume $M$ to be a compact, closed manifold
of a fixed topology. The integration (\ref{eq:Z_cont}) is taken over the 
space of the metric field $g_{\mu\nu}$ with an appropriate functional measure 
${\cal D} g$. 

  The groundwork for the path integral approach was laid by De~Witt~\cite{
DewittPI} and, subsequently, detailed prescriptions were developed by Gibbons, 
Hawking and Perry~\cite{GHP}. Although this approach has several
difficulties and unsolved problems, a number of highly suggestive results have
been obtained. Presumably, the dramatic successes of this approach are that, 
as mentioned in introduction, the foam picture of spacetime can be established 
semiclassically by calculating the contribution of the gravitational 
instantons~\cite{foamHawking}, and that the creation of particles near a 
Schwarzschild black hole can be related in a direct and simple manner to the
properties of the Euclidean Schwarzschild solution~\cite{HawkingRadtn}.
%
%
\begin{figure}[h]
\vspace{0.5cm}
\begin{center}
 \setlength{\unitlength}{1cm}
 \begin{picture}(15,4.3)(0,0)
 \put(0.0,0.1){\includegraphics[width=15cm]{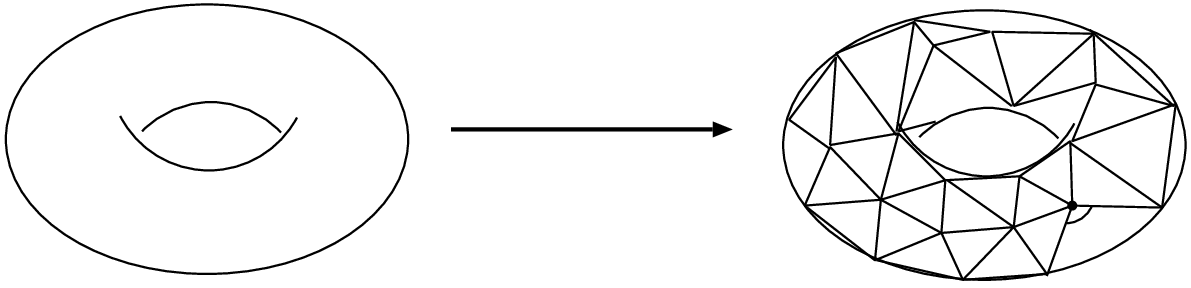}}
 \put(0.4,2.1){$g_{\mu\nu}$}
 \put(0.4,3.5){$M$}
 \put(6.52,2.3){\sl simplicial}
 \put(6.2,1.65){\sl decomposition}
 \put(10.3,3.5){$T$}
 \put(10.9,2.6){$l_1$}
 \put(10.5,2.1){$l_2$}
 \put(12.54,0.38){$l_i$}
 \put(14.0,1.6){$l_{N_1}$}
 \put(13.6,1.25){$h$}
 \put(13.6,0.75){$\epsilon_h$}
 \end{picture}
\end{center}
\vspace*{-0.8cm}
\caption{{\sl
An example of the simplicial decomposition. A smooth manifold $M$ 
with a metric $g$ is replaced by a simplicial manifold $T$ with a set of 
link-lengths $\{ l_1, l_2, \ldots, l_i, \ldots, l_{N_1} \}$. $\epsilon_h$ is 
the deficit angle around a hinge (vertex) $h$, corresponding to the scalar
curvature. In two dimensions the volume,
$A_h$, of the hinge $h$ is defined to be 1, while $A_h$ takes non-trivial 
values in higher dimensions.}}
\label{fig:simpMfdRep}
\vspace*{0.3cm}
\end{figure}

  Having shortly surveyed the continuum approach, let us now return to our main
interest. In Regge's lattice formulation of general relativity~\cite{Regge}, 
the continuum spacetime $M$ is replaced by a discretized space $T$ that 
consists of a finite number of $d$-simplices, as shown in 
Fig.~\ref{fig:simpMfdRep}. Such a discrete space $T$ is called a 
$d$-dimensional {\em simplicial manifold\/} or {\em triangulation\/} in lattice
gravity\footnote{Exactly speaking, the discrete counterpart of a Riemannian 
manifold $(M, g_{\mu\nu})$ is a {\em piecewise linear (PL) manifold}, which has
a certain metric structure that converges on the Riemannian structure in a 
continuum limit~\cite{CMS}. In other words, a simplicial manifold equipped with
the PL metric structure is assumed to be a (discrete) physical spacetime.}. 
First, we give a connectivity of vertices in $T$, which must satisfy the 
manifold condition that the space looks locally like a Euclidean space 
$\mbox{\boldmath $R$}^d$. Incidentally, the total number, $N_k$, of 
$k$-simplices also is determined $(k=0,\ts 1,\ts \ldots, \ts d)$. Next, we give
lengths $l_1,\ts l_2,\ldots,\ts l_{N_1}$ to all the links (1-simplices) in $T$.
As a result, the pair of the connectivity and the set of all the link-lengths 
$\{l_i\}$ determine the simplicial geometry on the discrete spacetime $T$.
This procedure is called the {\em simplicial decomposition}. An example 
of the procedure is depicted on Fig.~\ref{fig:simpMfdRep}, where a 
two-dimensional torus with a metric $g_{\mu\nu}$ is decomposed to a simplicial 
torus with the link-lengths $\{l_i\}$ by gluing triangles (2-simplices) along 
their boundary links (1-simplices). 

  On the simplicial manifold, in general, one can define geometric quantities 
(differential forms, curvatures, parallel transport, etc.) in a standard way. 
In particular, the Einstein-Hilbert action (\ref{eq:S_EH}) is replaced with 
the so-called {\em Regge action}, $S_{\rm Regge}$, on $T$~\cite{Regge}:
\begin{eqnarray}
S_{\rm Regge}[l,\ts T] = - \ts \frac{\ds 1}{\ds 16\pi G}
\nts \sum_{h \mbox{\sp :}\mbox{\sp hinges}} \nts 2A_h 
\epsilon_h + \ts {\mit\Lambda} \nts\nts\nts
\sum_{\sigma \mbox{\sp :} d \mbox{\sp -simplices}} \nts\nts
V_{\sigma} \es,
\label{eq:S_Regge}
\end{eqnarray}
where $A_{h}$ is the volume of a $(d-2)$-simplex ({\em hinge}) $h$ and 
$V_{\sigma}$ the volume of a $d$-simplex $\sigma$. The dimensionless quantity
$\epsilon_h$ in the right hand side of (\ref{eq:S_Regge}) is called the 
deficit angle around the hinge $h$, which plays the role of the local scalar 
curvature on $T$. The symbol $l$ denotes the set of all the link-lengths 
$\{l_1,\ts l_2,\ldots,\ts l_{N_1}\}$. 
\begin{figure}[h]
\vspace{0.6cm}
\begin{center}
 \setlength{\unitlength}{1cm}
 \begin{picture}(13.0,5.1)(0,0)
 \put(2.0,0.1){\includegraphics[width=9cm]{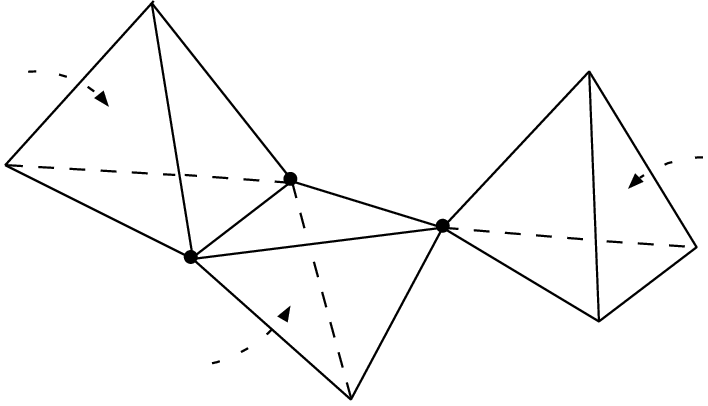}}
 \put(4.05,1.8){$c$} \put(5.8,3.1){$b$} \put(7.6,1.99){$a$}
 \put(5.3,5.1){\large {\sl simplicial complex} ${\cal C}$}
 \put(0.3,4.25){\sl manifold $T_1$} 
 \put(2.6,0.5){\sl manifold $T_2$} \put(11.1,3.2){\sl manifold $T_3$}
 \end{picture}
\end{center}
\vspace*{-0.8cm}
\caption
{{\sl
An example of a simplicial complex ${\cal C}$ that is {\it not\/} a manifold.
The complex ${\cal C}$ is a union of three simplicial submanifolds $T_i \es 
(i=1,2,3)$, and obviously violates the manifold condition at the junction 
parts, namely, at the vertex $a$ and along the link $bc$. The Regge action 
$S_{\rm Regge}[l,\ts T_i]$ is well-defined and finite on each $T_i$. 
Furthermore, the Regge action $S_{\rm Regge}[l,\ts {\cal C}]$ on ${\cal C}$ is
also defined uniquely as the sum of the actions 
$\sum_i S_{\rm Regge}[l,\ts T_i]$.
}}
\label{fig:sim_comp}
\vspace*{0.3cm}
\end{figure}

  Interestingly, the Regge action (\ref{eq:S_Regge}) is well-defined even on a 
degenerate configuration that is {\em not\/} a simplicial manifold but a
{\em simplicial complex}. The crucial difference between the simplicial 
manifold and the simplicial complex is that the former satisfies the manifold 
condition that the space looks locally like a Euclidean space 
$\mbox{\boldmath $R$}^d$, but the latter does {\em not necessarily}. 
Fig.~\ref{fig:sim_comp} shows a two-dimensional simplicial complex ${\cal C}$
which is not a manifold of definite dimension; the manifold condition is 
violated at the junction parts. However, the Regge action $S_{\rm Regge}
[l,\ts {\cal C}]$ on ${\cal C}$ is uniquely defined as the sum of the actions 
$\sum_i S_{\rm Regge}[l,\ts T_i]$ where $T_i \es (i\nts=\nts1,2,3)$ are the 
simplicial submanifolds of the complex ${\cal C}$. Unlike the case of the 
continuum action (\ref{eq:S_EH}), the finiteness of the Regge action 
(\ref{eq:S_Regge}) even on such a degenerate complex is a notable feature, 
particularly for describing the topology change of Euclidean spacetime in 
later section~\ref{sect:spacetimeFoam}.
%
%
\subsection{Quantum Regge calculus} 
\spc From now on let us concentrate on the problem of lattice quantization of 
gravity based on Regge calculus (RC). In quantum RC the discretized equivalent 
of the integration over the metrics (\ref{eq:Z_cont}) is implemented by 
{\em varying the link-lengths on a fixed triangulation }, assuming that 
patterns of triangulations are not dynamical at all; this is the reason why 
quantum RC is often called the {\em fixed triangulation (FT) approach}.
%
Accordingly, the partition function for quantum RC on a fixed
triangulation $T$ is given by
\begin{eqnarray}
Z_{\rm RC}[T] = \int d\mu_T[l] \ts
\exp \Bigl( - S_{\rm Regge}[l,\ts T] \Bigr) \es,
\label{eq:Z_RC}
\end{eqnarray}
where $d\mu_T[l]$ is a link-integration measure defined on $T$, including 
constraints imposed by the (higher-dimensional analogs of) triangle 
inequalities. Each length $l_i$ is integrated over a range $l_{\rm min} < l_i 
< l_{\rm max}$. In eq.~(\ref{eq:Z_RC}) we explicitly write down the 
`subscript' $T$ of the measure $d\mu_T[l]$ and the `variable' $T$ for the 
partition function $Z_{\rm RC}[T]$ in order to make it clear that they are 
defined on the triangulation $T$. This notation is useful for our formulation.

  The FT approach (\ref{eq:Z_RC}) has been actively applied to quantization of
gravity~\cite{HamRev,HWbasic} and actually has many appealing aspects. 
However, this approach has some subtle issues. One of them is how to define the
discretized measure $d\mu_T[l]$ that corresponds to the continuum measure 
${\cal D} g$ in eq.~(\ref{eq:Z_cont}) over the so-called 
superspace~\cite{DeWittSprSp} of the Riemannian metrics on $M$. In fact, one 
has to gauge away the diffeomorphism group $\mbox{Diff}_M$ that is the gauge 
freedom of general relativity. In the continuum a possibility is to start from 
the supermetric formulation~\cite{DeWittSprMtrc} that defines a gauge-invariant
measure over the superspace and then to gauge away $\mbox{Diff}_M$ by using 
one's favorite gauge.

  Here one would ask a simple question: What is the lattice counterpart of 
the diffeomorphism group $\mbox{Diff}_M$?  Though we have no clear answer to
the question, the notion of ``diffeomorphism'' on the lattice has been used at 
two different levels~\cite{progRC}. One definition of the symmetry, which is 
often adopted by those studying Regge calculus in classical relativity, is 
called ``invariance of the geometry'' in which transformations of link-lengths 
leave the {\em geometry\/} invariant. A proper implementation of the definition
is to require that all local curvatures (and hence all deficit angles) should 
be unchanged under the transformations. But this requirement is too strong to 
satisfy in the FT approach\footnote{The only exception is flat space where an 
infinite number of choices of the link-lengths will generally correspond to the
same flat geometry even in the FT approach.}~\cite{HW2}.

  Another definition, favored by those wishing to use results from lattice 
gauge theories, is referred to as ``invariance of the action'' in which 
transformations of link-lengths leave the {\em action\/} invariant. It is 
possible even in quantum RC to imagine that changes in the link-lengths which 
could increase deficit angles in one region and compensatingly decrease them in
another would produce no overall change in the Regge action (\ref{eq:S_Regge}).
However, the symmetry in this sense holds approximately at most
in (almost) flat space~\cite{RW1,HW2,HW3}; still less it would be possible
to find in curved space any exact symmetry that can be interpreted as 
the lattice diffeomorphism in this approach.

\begin{figure}[h]
\vspace{0.2cm}
\begin{center}
 \setlength{\unitlength}{1cm}
 \begin{picture}(14,2.0)(0,0)
 \put(0.0,0.0){\includegraphics[width=14cm]{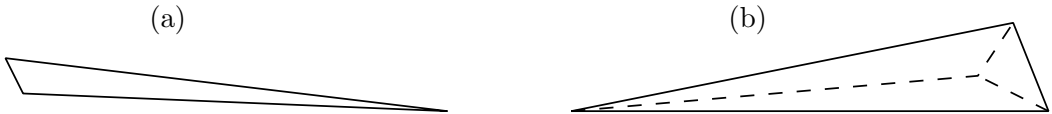}}
 \put(2.0,1.2){(a)}  \put(9.7,1.2){(b)} 
 \end{picture}
\end{center}
\vspace*{-0.5cm}
\caption{
Singular configurations arising in the FT approach. 
(a) A singular 2-simplex. (b) A singular 3-simplex.
}
\label{fig:simpSnglr}
\end{figure}

  The other problem is that there can be generally singular configurations 
including such simplices that have very short and very long links at the same
time, as shown in Fig.~\ref{fig:simpSnglr}. Though one usually deals with this 
problem by putting upper and lower bounds on the lengths $l_i$, such 
configurations are unavoidable to obtain a metric with very high curvature out 
of the flat metric. It is still unclear whether one can explore regions in the 
space of metrics where the metric is very singular and fairly different from 
the typical metric on the reference simplicial space that we have chosen at the
beginning. It will cause a large difficulty in studying the strong-coupling 
region of lattice quantum gravity, where configurations with very high 
curvatures are expected to emerge frequently.
%
%
%
%
\subsection{Dynamical triangulations}
\spc In the dynamical triangulation (DT) approach~\cite{Weingarten,DT}, which 
is the alternative to the FT one, it is assumed that none of the link-lengths 
in each triangulation are dynamical and all of them can be fixed to a single 
lattice spacing $a$. Accordingly, the simplicial manifold $T$ considered in DT 
consists of equilateral $d$-simplices. The Regge action (\ref{eq:S_Regge}) on 
such a equilateral triangulation $T$ becomes the following simple form:
\begin{eqnarray}
S_{\mbox{\sp DT}}[T] = -\kappa_0 N_0 + \kappa_d N_d \es,
\label{eq:S_DT}
\end{eqnarray}
where $\kappa_0$ and $\kappa_d$ are constants related to $G,\ts {\mit\Lambda}$ 
and $a$. $N_0$ and $N_d$ stand for the number of $0$-simplices and that of
$d$-simplices on $T$. Instead of the link-lengths, we take {\em varying 
connectivity of vertices\/} as dynamical.
Hence, the path integral for DT is defined by the sum over all possible 
patterns of triangulations that consist of equilateral $d$-simplices:
\begin{eqnarray}
Z_{\rm DT} 
\es = \nts\sum_{T : \mbox{\sp triangulations}} \nts\nts
\exp \Bigl( - S_{\rm DT}[T] \Bigr) \ts.
\label{eq:Z_DT}
\end{eqnarray}
Here we fix the topology of all the triangulations $T$.
%
%
\begin{figure}
\begin{center}
 \setlength{\unitlength}{1cm}
 \begin{picture}(12,6.0)(0,0)
 \put(0.0,0.0){\includegraphics[width=12cm]{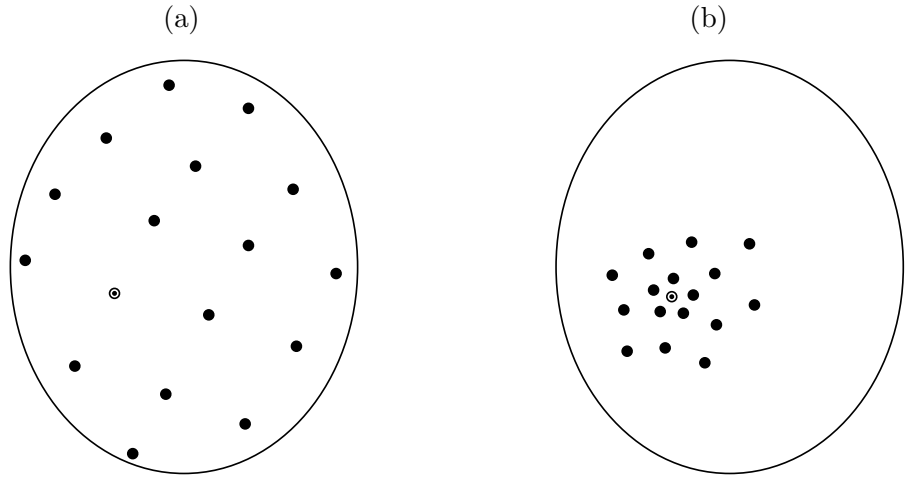}}
 \put(2.1,6.0){(a)}  \put(9.1,6.0){(b)} 
 \end{picture}
\end{center}
\vspace*{-0.8cm}
\caption{{\sl
Sketch of the space of all Riemannian metrics. Each point means a simplicial 
geometry. The circles with dot represent the reference (flat) simplicial space 
at the beginning. {\rm (a)} ``Regular'' distribution arising in the DT 
approach. {\rm (b)} ``Localized '' distribution around the reference metric 
arising in the FT approach.
}}
\label{fig:dstrnMtrcs}
\end{figure}
\vspace*{0.3cm}

  In what follows, we make a brief comparison between FT (\ref{eq:Z_RC}) and 
DT (\ref{eq:Z_DT}) to clarify the differences between them. For any 
Riemannian manifold non-singular enough to start as the reference at the 
beginning, one can construct a sequence of simplicial manifolds. We expect that
in DT the discrete set of simplicial spaces is regularly distributed in the 
space of all the Riemannian metrics~\cite{RZ}, as shown in 
Fig.~\ref{fig:dstrnMtrcs} (a). In contrast, simplicial spaces arising in FT are
localized in the neighborhood of the smooth metric on the regular lattice that 
has been chosen at the beginning (see Fig.~\ref{fig:dstrnMtrcs} (b)); such a 
localization will make the lattice configurations less accessible to the 
strong-coupling regions that might be far from the reference metric, and 
prevent ``diffeomorphism-invariance'' from holding. In this respect, 
DT has the advantage over FT.
Moreover, DT has the natural UV cutoff $a$ of the theory.

  Nevertheless, the DT approach makes some issues more difficult. Actually, we 
have completely lost diffeomorphism-invariance on the lattice, because one 
cannot deform smoothly (even approximately) a simplicial lattice into another 
in this approach. Some argue that in DT the permutation group $S_{N_0}$ acting 
on $N_0$ vertices is expected to reproduce the diffeomorphism group acting on 
a smooth manifold in a continuum limit\footnote{In the IIB matrix model, which 
is a candidate for a constructive definition of superstring theory, the same 
scenario is also discussed~\cite{IIBIK}.}, although no exact proof has been 
known yet. One has to check whether diffeomorphism-invariance is recovered in 
the continuum limit\footnote{In the case of two-dimensional quantum gravity, 
however, it has been verified that results obtained in DT~\cite{DT} coincide 
with the predictions from CFT~\cite{DK} and diffeomorphism-invariance is 
recovered in two dimensions.}, if it exists. In addition, it is not clear how 
to define the classical continuum limit of DT, since this approach gives no 
field equation corresponding to Einstein's equation. As a consequence, this 
difficulty makes it fairly intractable to study (quantum) black hole physics
in DT.
\vspace*{0.3cm}

%
%
   Given reasonable geometric and symmetric properties, universality in quantum
field theory will generally assure the same results for physical observables 
in a certain continuum limit, even though the details of UV behaviors of two
field theories are different from each other. This is almost the case with 
lattice field theories, where physical results are expected to be independent 
of the specific details of the UV cutoff and the specific forms of the 
lattice actions as long as theories of interest preserve the underlying 
symmetries. 
In the case of lattice gravity, however, such a naive expectation from 
universality will not necessarily hold, because the two approaches have 
different symmetric properties as discussed above. 
Exactly speaking, it is a drawback to the 
lattice regularizations of gravity that they lack diffeomorphism-invariance 
even at the classical level, in contrast to other lattice field 
theories where gauge symmetries are (classically) exact on the finite lattice. 
Attempts to overcome the drawback strongly motivate us to study the hybrid 
model of quantum RC and DT, namely, dynamical Regge calculus, from now on.
\clearpage
%

%
\section{Dynamical Regge calculus}\label{sect:defDRC}
\spc In this section we first give the definition of our hybrid model as a 
simultaneous implementation of quantum RC and DT and discuss its fundamental 
properties. We discuss possible behaviors of the entropy\footnote{Here the 
entropy means the total number of possible triangulations in the model and, 
therefore, it is {\em not\/} the black hole entropy. Do not be confused with 
the same term `entropy'.} of the model, which is related directly to the 
well-definedness of the model. Then, we formulate elementary local moves 
necessary to carry out numerical simulations. Finally, we will give an 
extension of the hybrid model including degenerate configurations and apply it 
to a topology-changing process of Euclidean spacetime on the lattice.
%
\subsection{Definition of the hybrid model}
\begin{figure}[h]
\vspace{0.5cm}
\begin{center}
 \setlength{\unitlength}{1cm}
 \begin{picture}(13.2,6.0)(0,0)
 \put(2.1,0.0){\includegraphics[width=9cm]{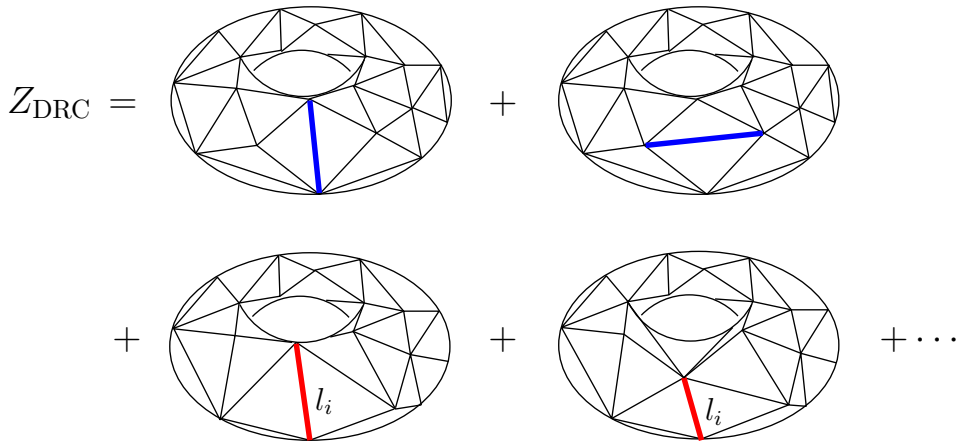}}
 \put(0.0,4.45){\Large $Z_{\rm DRC} \ts = $} \put(6.4,4.45){\Large $+$}
 \put(1.4,1.3){\Large $+$} \put(6.4,1.3){\Large $+$}
 \put(11.6,1.3){\Large $+ \cdots$}
 \put(4.1,0.5){\large $l_i$} \put(9.28,0.36){\large $l_i$}
 \end{picture}
\end{center}
\vspace{-0.8cm}
\caption
{{\sl 
Lattice configurations appearing in the sum {\rm (\ref{eq:Z_DRC})}.
The first two configurations have different patterns of connectivity, arising 
from the triangulation-sum $\sum_T$. The next two have the same connectivity
but contain different lengths of the $i$-th link, arising from the 
link-integration $\int d\mu_T[l]$.
}}
\label{fig:partSumDRC}
\vspace{0.3cm}
\end{figure}

\spc The partition function, $Z_{\rm DRC}$, of {\em dynamical Regge calculus\/}
(DRC) is defined by performing {\em at once\/} the link-integration 
(\ref{eq:Z_RC}) and the triangulation-sum (\ref{eq:Z_DT}):
\begin{eqnarray}
Z_{\rm DRC} 
\ts = \sum_{T{\rm :triangulations}} \nts\int d\mu_T[l] \ts
\exp \Bigl( - S_{\rm Regge}[l,\ts T] \Bigr) 
\ts = \ts\sum_{T{\rm :triangulations}} \nts\nts Z_{\rm RC}[T] \es,
\label{eq:Z_DRC}
\end{eqnarray}
where each link-length $l_i$ is integrated over the range $l_{\rm min} < l_i < 
l_{\rm max}$. The sum $\sum_{T{\rm :triangulations}}\int d\mu_T[l]$ means that 
any pattern of triangulation (connectivity of vertices) $T$ with various sets 
of the link-lengths is included so long as it satisfies both the manifold 
condition and the triangle inequalities (see Fig.~\ref{fig:partSumDRC}). 
The reason for the name of dynamical Regge calculus is that Regge calculus 
$Z_{\rm RC}[T]$ on each triangulation $T$ is thought
of as if a `dynamical variable' in the partition sum (\ref{eq:Z_DRC}). 
%
%
%
%
%

%
%
   A few remarks are in order. First, the behavior of the entropy 
of the hybrid model (\ref{eq:Z_DRC}) is essential to its well-definedness 
as a statistical system. A natural generalization from DT leads to the 
following definition of the entropy, $W(N_d)$, of the model (\ref{eq:Z_DRC}):
\begin{eqnarray}
W(N_d) \equiv \sum_{T : N_d \ts \mbox{\sp fixed}}
\int d\mu_T[l] \es,
\label{def:entropy_DRC}
\end{eqnarray}
where the sum $\sum_{T : N_d \ts \mbox{\sp fixed}}$ is taken over 
triangulations with the number of $d$-simplices $N_d$ fixed. 
It depends strongly on the behavior of the entropy (\ref{def:entropy_DRC}) 
whether the model (\ref{eq:Z_DRC}) is well-defined statistical-mechanically 
or not. If we set the coupling constant $G = \infty$ for simplicity,
the following inequality holds:
\begin{eqnarray}
e^{-S_{\rm Regge} [l,\vts T]} < 
e^{-{\mit \Lambda} \vts l_{\rm min}^d N_d} \es,
\label{eq:Z_DRC_ineq}
\end{eqnarray}
where the minimal length $l_{\rm min}$ plays the role of the UV cutoff. Thus, 
for the partition function $\tilde{Z}_{\mbox{\sp DRC}}(N_d)$ with $N_d$ 
fixed, one easily obtains the following inequality:
\begin{eqnarray}
\tilde{Z}_{\rm DRC}(N_d) 
&\equiv& \nts\nts \sum_{T : N_d \ts \mbox{\sp fixed}} 
\int d\mu_T[l] \es \exp 
\Bigl( - S_{\rm Regge}[l,\vts T] \Bigr) \nonumber \\
&  <   & W(N_d) \es e^{-{\mit \Lambda} \vts l_{\rm min}^d N_d} \es.
\label{eq:W_def}
\end{eqnarray}
If an exponential bound for the entropy $W(N_d)$ holds
\begin{eqnarray}
W(N_d) \ts < \es const. \times e^{{\mit \Lambda}_c l_{\rm min}^d N_d} 
\quad\quad ({\mit \Lambda}_c \ts \mbox{; a positive constant}) \es,
\label{eq:W_bound}
\end{eqnarray}
then DRC (\ref{eq:Z_DRC}) is well-defined at least as a statistical system. 
In this case it is a possibility that one can take a continuum limit by 
fine-tuning the cosmological constant ${\mit \Lambda}$. However, it is very 
hard to give any analytic proof for the exponential bound (\ref{eq:W_bound})
because the integration measure $d\mu_T[l]$ includes the triangle inequalities 
which are too intractable to calculate analytically. Instead, we will later see
a piece of numerical evidence that the bound (\ref{eq:W_bound}) holds in case 
that we use the scale-invariant measure $\prod_i dl_i / l_i$.

  Second, we comment on a direct relation between DT (\ref{eq:Z_DT}) and 
DRC (\ref{eq:Z_DRC}). If one chooses the $\delta$-function measure
\begin{eqnarray}
d\mu_T[l] = \prod_i dl_i \ts \delta (l_i - a) \es,
\label{def:deltaFuncMeasure}
\end{eqnarray}
then DRC (\ref{eq:Z_DRC}) becomes the same as DT (\ref{eq:Z_DT}) where all the 
lengths are fixed to the single lattice spacing $a$ and the triangle 
inequalities are automatically satisfied. In other words, DT is equivalent to 
DRC with the $\delta$-function measure (\ref{def:deltaFuncMeasure}). As is well
known with matrix models~\cite{GM}, such a choice (\ref{def:deltaFuncMeasure}) 
gives a constructive definition of two-dimensional quantum gravity, even though
the measure (\ref{def:deltaFuncMeasure}) breaks explicitly the lattice 
diffeomorphism-invariance. Although the problem of the lattice symmetry is not 
dealt with here, we will observe in section~\ref{sect:genCovLat} that it is a 
difficult task to construct a `diffeomorphism-invariant' measure
on the lattice.
%
%
%
%
\subsection{Hybrid $(p,\ts q)$ moves in DRC}\label{sect:hybridMoves}
\spc Here we construct local, ergodic moves, i.e., local changes of 
triangulations, though they never change the
topology of the simplicial manifolds; such elementary moves are a necessary 
ingredient to carry out numerical studies of DRC (\ref{eq:Z_DRC}).
The Monte-Carlo method is applicable to DRC\cite{DRC,DRC2}, as well as it has 
been to both quantum RC~\cite{RW1,HW3} and DT~\cite{DT}. Remember that in DT 
the so-called $(p,\ts q)$ moves\footnote{In ref.~\cite{AGV}, these moves are 
called the $(k,\ts l)$ moves. To avoid confusion, however, we use here the term
of the $(p,\ts q)$ moves instead, since we use the symbol $l$ to denote (a set 
of) link-lengths.} are used and they are ergodic in the class of triangulations
of fixed topology~\cite{AGV}. In what follows, we give an extension of the 
$(p,\ts q)$ moves to invoke quantum RC in addition to DT according to 
DRC (\ref{eq:Z_DRC}). 
%
%
\begin{figure}[h]
\vspace*{0.5cm}
\begin{center}
 \setlength{\unitlength}{1cm}
 \begin{picture}(14.0,9.4)(0,0)
 \put(0.7,0.3){\includegraphics[width=13cm]{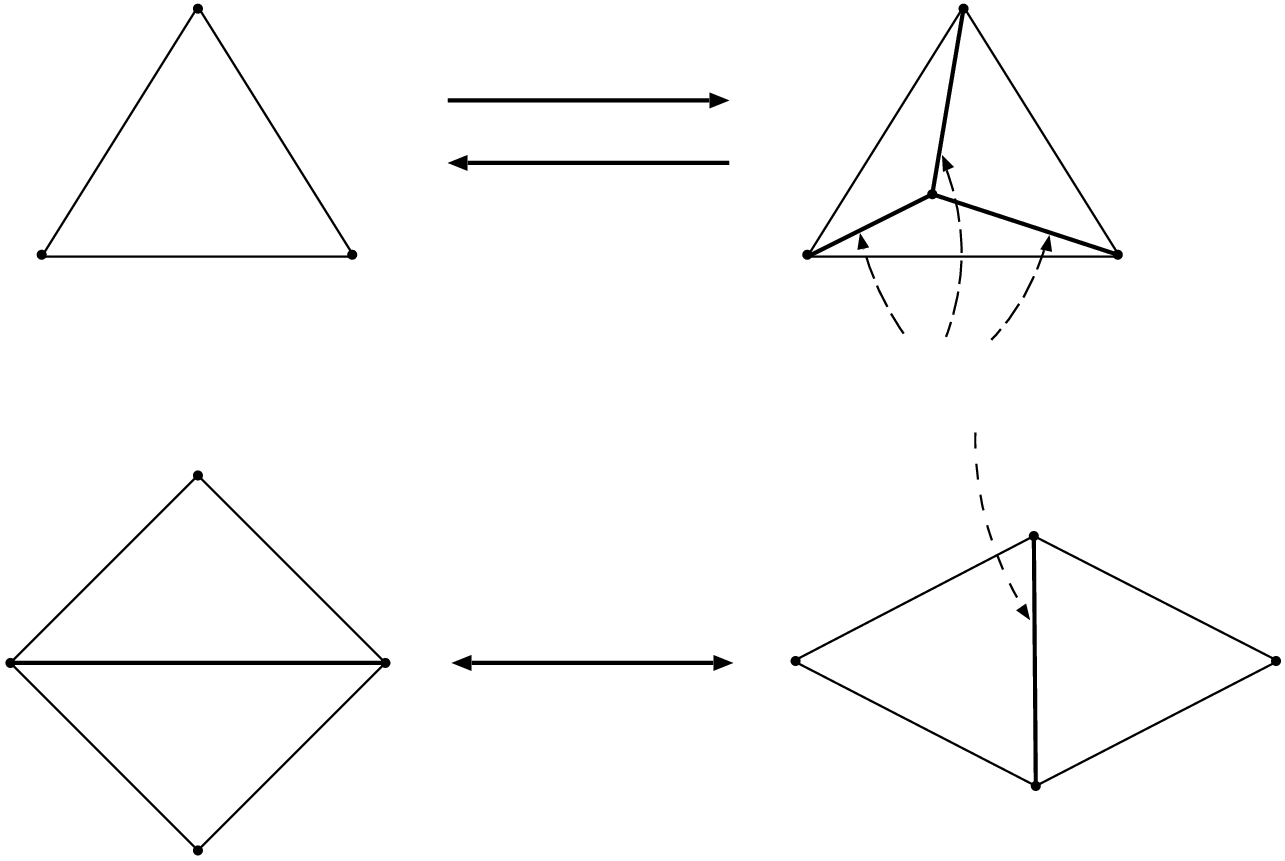}}
 \put(1.0,9.1){(a)}
 \put(2.55,9.15){$v_0$} \put(0.9,6.0){$v_1$} \put(4.2,6.0){$v_2$}
 \put(5.0,8.2){\sl hybrid $(1,\ts 3)$ move}
 \put(5.0,6.9){\sl hybrid $(3,\ts 1)$ move}
 \put(10.25,9.15){$v_0$} \put(8.65,6.0){$v_1$} \put(11.85,6.0){$v_2$}
 \put(10.4,7.7){$l_0$} \put(9.9,7.15){$v$} 
 \put(9.4,6.9){$l_1$} \put(11.0,6.9){$l_2$}
 \put(9.0,5.0){\sl arbitrary lengths} 
 \put(1.0,4.4){(b)}
 \put(2.6,4.4){$v_0$} \put(2.6,0.0){$v_2$}
 \put(0.45,1.95){$v_1$} \put(4.6,1.95){$v_3$} \put(2.7,2.5){$l$}
 \put(5.1,2.65){\sl hybrid $(2,\ts 2)$ move}
 \put(8.5,1.96){$v_1$} \put(13.5,1.96){$v_3$} \put(11.3,2.3){$\tilde{l}$}
 \put(11.05,3.88){$v_0$} \put(11.08,0.6){$v_2$}
 \end{picture}
\end{center}
\vspace*{-0.8cm}
\caption
{{\sl 
Hybrid $(p,\ts q)$ moves in two dimensions. {\rm (a)} A hybrid $(1,\ts 3)$ 
move, where the triangle $v_0v_1v_2$ is divided into three triangles $vv_0v_1,
\ts vv_1v_2$ and $vv_2v_0$. Then, inserted links $vv_0,\ts vv_1$ and $vv_2$ can
take arbitrary lengths so long as triangle inequalities hold. Its inverse move,
called the hybrid $(3,\ts 1)$ move, is also depicted. {\rm (b)} A hybrid 
$(2,\ts 2)$ move acting on two triangles $v_0v_1v_3 + v_1v_2v_3$. After 
deleting the link $v_1v_3$ of length $l$, a new link $v_0v_2$ of length 
$\tilde{l}$ is inserted. The length $\tilde{l}$ is also arbitrary unless 
triangle inequalities are broken.
}}
\label{fig:ex_kl_moves_2}
\vspace{0.3cm}
\end{figure}

  In Fig.~\ref{fig:ex_kl_moves_2}, we show an example of the extended moves in 
two-dimensions; in Fig.~\ref{fig:ex_kl_moves_2}~(a), a triangle $v_0v_1v_2$ is 
divided to three new triangles $vv_0v_1,\ts vv_1v_2$ and $vv_2v_0$ by 
inserting a vertex $v$ and three new links $vv_0,\ts vv_1$, $vv_2$.
The link-lengths are arbitrary as long as triangle inequalities hold.
The deficit angles around the vertices $v_0,\ts v_1,\ts v_2$ and $v$ can change
continuously in DRC as well as in quantum RC\footnote{Indeed, we observe soon 
later that the link-integration of quantum RC is automatically included in such
moves.}. The inverse move is defined straightforwardly, as shown in 
Fig.~\ref{fig:ex_kl_moves_2}~(a). We call the two moves of 
Fig.~\ref{fig:ex_kl_moves_2}~(a) the {\em hybrid $(1,\ts 3)$ move\/} and the 
{\em hybrid $(3,\ts 1)$ move}. Similarly, one can extend the $(2,\ts 2)$ move 
of DT to a hybrid one of DRC, as depicted on Fig.~\ref{fig:ex_kl_moves_2}~(b). 
The link $v_1v_3$ of length $l$ shared by two triangles $v_0v_1v_3$ and
$v_1v_2v_3$ is flipped to a new link $v_0v_2$ of length $\tilde{l}$;
$\tilde{l}$ also is arbitrary unless triangle inequalities are broken and, 
hence, the deficit angles around the vertices $v_0,\ts v_1,\ts v_2$ and $v_3$ 
can take continuous values. We call the extended move in 
Fig.~\ref{fig:ex_kl_moves_2}~(b) the {\em hybrid $(2,\ts 2)$ move}.

\begin{figure}[h]
\vspace{0.2cm}
\begin{center}
 \setlength{\unitlength}{1cm}
 \begin{picture}(12,10.3)(0,0)
 \put(0.5,0.5){\includegraphics[width=11cm]{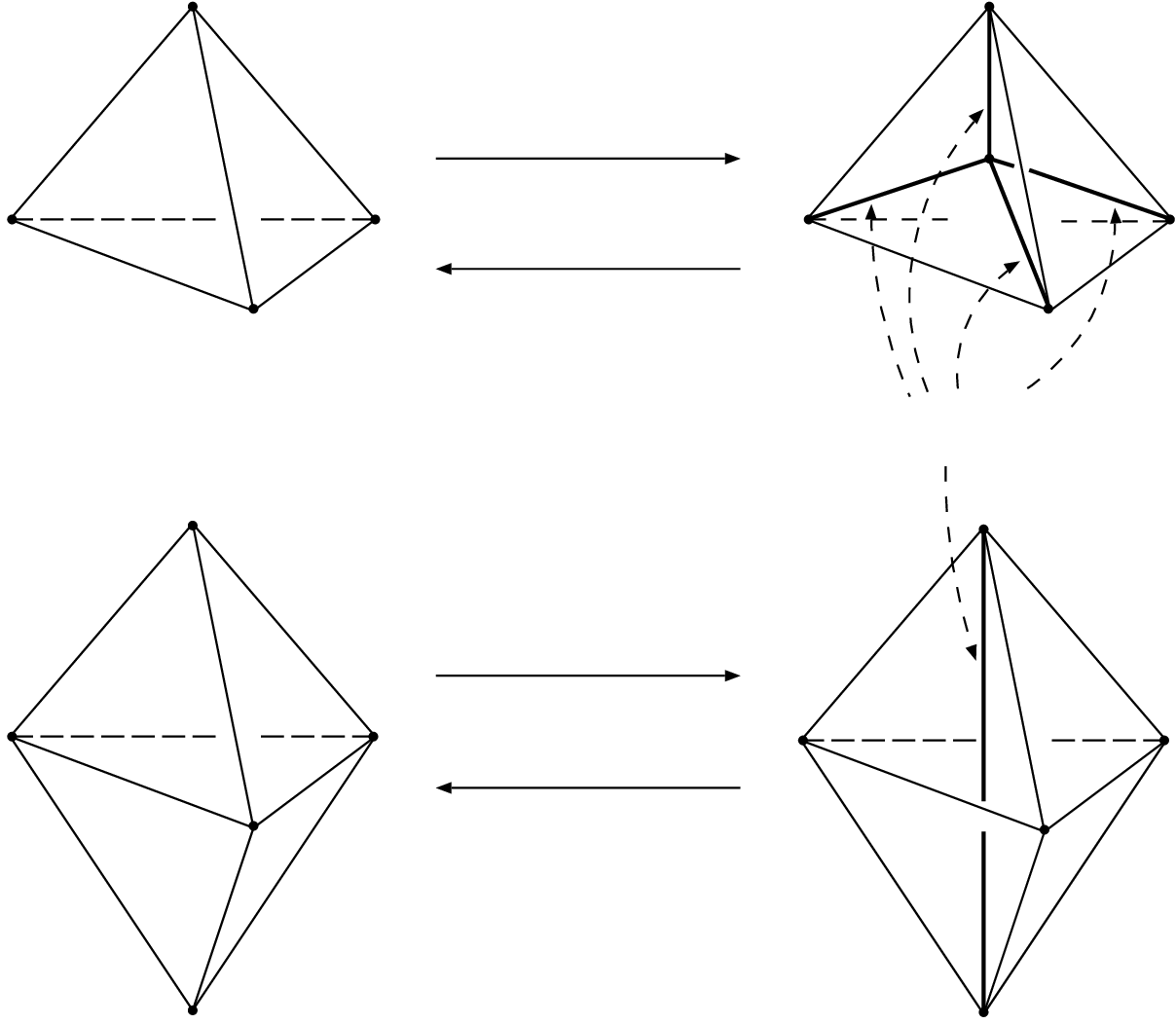}}
 \put(0.2,9.8){(a)}
 \put(2.2,10.2){$v_0$} \put(0.18,8.0){$v_1$}
 \put(2.6,6.9){$v_2$} \put(4.18,8.0){$v_3$}
 \put(4.4,8.9){\sl hybrid $(1,\ts 4)$ move}
 \put(4.4,7.0){\sl hybrid $(4,\ts 1)$ move}
 \put(9.6,10.2){$v_0$} \put(7.6,8.0){$v_1$} \put(9.5,8.2){$v$}
 \put(10,6.9){$v_2$} \put(11.6,8.0){$v_3$} \put(9.48,9.25){$l_0$}
 \put(8.7,8.4){$l_1$} \put(9.7,7.8){$l_2$} \put(10.6,8.45){$l_3$}
 \put(7.8,6.0){\sl arbitrary lengths} 
 \put(0.2,5.2){(b)}
 \put(2.2,5.35){$v_0$} \put(2.2,0.25){$v_4$}
 \put(0.18,3.1){$v_1$} \put(2.37,2.15){$v_2$} \put(4.13,3.1){$v_3$}
 \put(4.4,4.0){\sl hybrid $(2,\ts 3)$ move}
 \put(4.4,2.16){\sl hybrid $(3,\ts 2)$ move}
 \put(9.6,5.35){$v_0$} \put(9.6,0.25){$v_4$} \put(9.5,3.4){$l$}
 \put(7.58,3.1){$v_1$} \put(9.77,2.15){$v_2$} \put(11.53,3.1){$v_3$}
 \end{picture}
\end{center}
\vspace*{-0.8cm}
\caption {{\sl
Hybrid $(p,\ts q)$ moves in three dimensions. {\rm (a)} A hybrid $(1,\ts 4)$ 
move, where a new vertex $v$ is put into the 3-simplex $vv_0v_1v_2v_3$ and four
links $vv_0,\ts vv_1,\ts vv_2$ and $vv_3$ with arbitrary lengths $l_0,\ts l_1,
\ts l_2$ and $l_3$ are inserted unless triangle inequalities are broken. Its 
inverse move is also depicted. {\rm (b)} A hybrid $(2,\ts 3)$ move acting
on two 3-simplices $v_0v_1v_2v_3 + v_1v_2v_3v_4$. Instead of the triangle 
$v_1v_2v_3$, the new link $v_0v_4$ is inserted. Length $l$ of $v_0v_4$ is 
arbitrary so long as triangle inequalities hold. Its inverse moves, called 
the hybrid $(3,\ts 2)$ move, is also shown.
}}
\label{fig:ex_kl_moves_3}
\vspace{0.5cm}
\end{figure}

%
%
  It is straightforward to generalize the two-dimensional hybrid 
$(p,\ts q)$ moves to higher-dimensional ones, where $p + q = d + 2 \es 
(d = 3,\ts 4)$. For example, Fig.~\ref{fig:ex_kl_moves_3} shows
three-dimensional hybrid $(p,\ts q)$ moves. 
%
%

\begin{figure}[h]
\vspace{0.0cm}
\begin{center}
 \setlength{\unitlength}{1cm}
 \begin{picture}(16,3.1)(0,0)
 \put(0.5,0.2){\includegraphics[width=15cm]{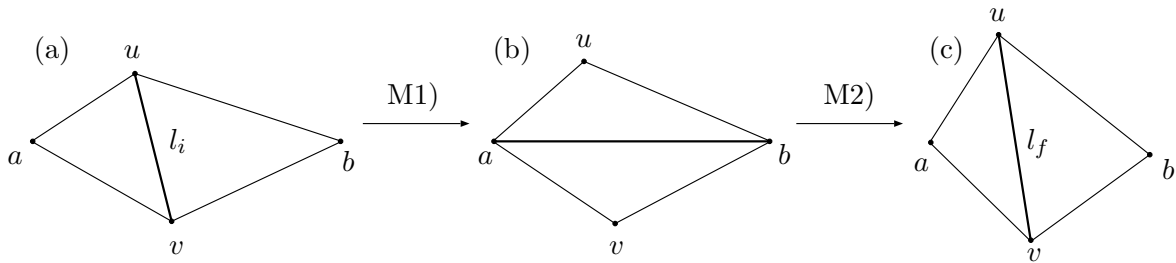}}
 \put(0.6,2.7){(a)}
 \put(1.8,2.7){$u$} \put(2.4,0.1){$v$}
 \put(0.25,1.3){$a$} \put(4.7,1.2){$b$} \put(2.4,1.5){$l_i$}
 \put(5.3,2.1){M1)}
 \put(6.7,2.7){(b)}
 \put(7.8,2.85){$u$} \put(8.25,0.1){$v$}
 \put(6.5,1.3){$a$} \put(10.5,1.3){$b$}
 \put(11.1,2.1){M2)}
 \put(12.5,2.7){(c)}
 \put(13.3,3.2){$u$} \put(13.8,0.0){$v$}
 \put(12.3,1.2){$a$} \put(15.6,1.1){$b$} \put(13.8,1.5){$l_f$}
 \end{picture}
\end{center}
\vspace*{-0.8cm}
\caption
{{\sl 
An example of the realization of a link-integration by successive applications 
of the hybrid $(p,\ts q)$ moves in two dimensions. Through the first hybrid 
$(2,\ts 2)$ move {\rm M1)}, the link $uv$ of length $l_i$ is flipped to a 
link $ab$. The next hybrid $(2,\ts 2)$ move {\rm M2)} changes the link $ab$ 
back to the link $uv$ of length $l_f$. As a result, this process realizes 
a link-integration of quantum RC only by the hybrid moves.
}}
\label{fig:Regge_moves_2}
\vspace{0.5cm}
\end{figure}

  The hybrid $(p,\ts q)$ moves described above are expected to be ergodic in 
the sense that all the lattice configurations can be generated only by 
the hybrid $(p,\ts q)$ moves.
Moreover, in Fig.~\ref{fig:Regge_moves_2} we explain how the hybrid moves 
implement not only DT but also quantum RC .
A link-integration will be done by successive applications of the hybrid
moves as follows:
\begin{enumerate}
\setlength{\itemsep}{-2pt}
\item[M1)] On a two-dimensional simplicial lattice we pick up a link $uv$ 
of length $l_i$ (see Fig~\ref{fig:Regge_moves_2}~(a)). Then, we apply a hybrid 
$(2,\ts 2)$ move to the link $uv$, and a new link $ab$ is created (see 
Fig~\ref{fig:Regge_moves_2}~(b)).
\item[M2)] Another hybrid $(2,\ts 2)$ move is applied to the link $ab$, 
getting back to the same connectivity as the initial configuration
(see Fig~\ref{fig:Regge_moves_2}~(c)). The only
difference between the initial and final configurations is that the length 
$l_i$ of the link $uv$ is replaced with the different one $l_f$.
\end{enumerate}
Through the processes M1) and M2), the length of the link $uv$ has changed 
from the initial $l_i$ to the final one $l_f$ according to the proper 
Boltzmann weight. Hence, this is a typical example of the realization of 
quantum RC only by a combinations of the hybrid moves.
%
%

  One would guess that in higher dimensions similar combinations 
of the hybrid moves can incorporate link-integrations of quantum RC into DRC.
This is really the case with our hybrid model.
In addition to the link-integration discussed above, the hybrid $(p,\ts q)$ 
moves are ergodic in the sense that all the patterns of triangulations can be 
generated by the hybrid moves owing to the established ergodicity of the usual
$(p,\ts q)$ moves in DT~\cite{AGV}.
As a consequence, these properties complete the ergodicity of the hybrid 
$(p,\ts q)$ moves in DRC. 
%
%
\subsection{Description of topology change in DRC}\label{sect:spacetimeFoam}
\spc As an application of the hybrid model, we try to describe the
topology-changing processes of Euclidean spacetime on the lattice. 
For the purpose, we need to extend our model (\ref{eq:Z_DRC}) in order to deal 
with degenerate configurations that appear in the topology change, as will be
discussed below.

  Over forty years ago, Wheeler pointed out~\cite{spcTmFoam} that the 
Einstein-Hilbert action (\ref{eq:S_EH}) allows large fluctuations of the metric
and even of the topology of spacetime manifolds on scales of order of the 
Planck length. This is due to the fact that the action for the 
gravitational field (\ref{eq:S_EH}) is not scale invariant, unlike that for the
Yang-Mills fields. Hence, a large fluctuation of the metric over a short length
scale does not have a very large value of the action (\ref{eq:S_EH}) and so is 
not highly damped in the path integral (\ref{eq:Z_cont}). The resultant quantum
spacetime with the large fluctuations of both the metric and the topology
is called the {\em spacetime foam}.
  Subsequently, Hawking calculated semiclassically the path integral for the 
spacetime foam by summing up ``gravitational instantons\footnote{For a detailed
explanation of the gravitational instantons, see ref.~~\cite{EGH}.}'' with 
various Euler numbers $\chi$ in four dimensions~\cite{foamHawking}. 
Physically, the ``gravitational instantons'' describe the topology-change of 
Euclidean spacetime. Furthermore, Hawking discussed a dynamical realization of 
such a topology-changing process by using quantum Regge calculus;
a metric can change topology without increasing the lattice action by 
more than an arbitrary small amount~\cite{foamHawking}. 
In what follows, we explain the close relation between Hawking's idea and 
our hybrid model.

  We first define the partition function, $Z_{\rm EDRC}$, 
of the extended hybrid model:
\begin{eqnarray}
Z_{\rm EDRC}
\ts = \sum_{{\cal C}{\rm :complexes}} \nts\int d\mu_{\cal C}[l] \ts
\exp \Bigl( - S_{\rm Regge}[l,\ts {\cal C}] \Bigr) 
\ts = \sum_{{\cal C}{\rm :complexes}} \nts\nts Z_{\rm RC}[{\cal C}] \es,
\label{eq:Z_DRC_ext}
\end{eqnarray}
where the sum $\sum_{{\cal C}{\rm :complexes}}\int d\mu_{\cal C}[l]$ admits
degenerate simplicial complexes (see Fig.~\ref{fig:sim_comp}) in addition to 
simplicial manifolds. Furthermore, manifolds (or complexes) of various 
topologies are contained in the extended sum (\ref{eq:Z_DRC_ext}) in order to 
describe topology-changing processes dynamically.

\begin{figure}
\vspace*{-0.1cm}
\begin{center}
\setlength{\unitlength}{1cm}
\begin{picture}(16,18.6)(0,0)
\put(0.6,0.1){\includegraphics[height=18.0cm]{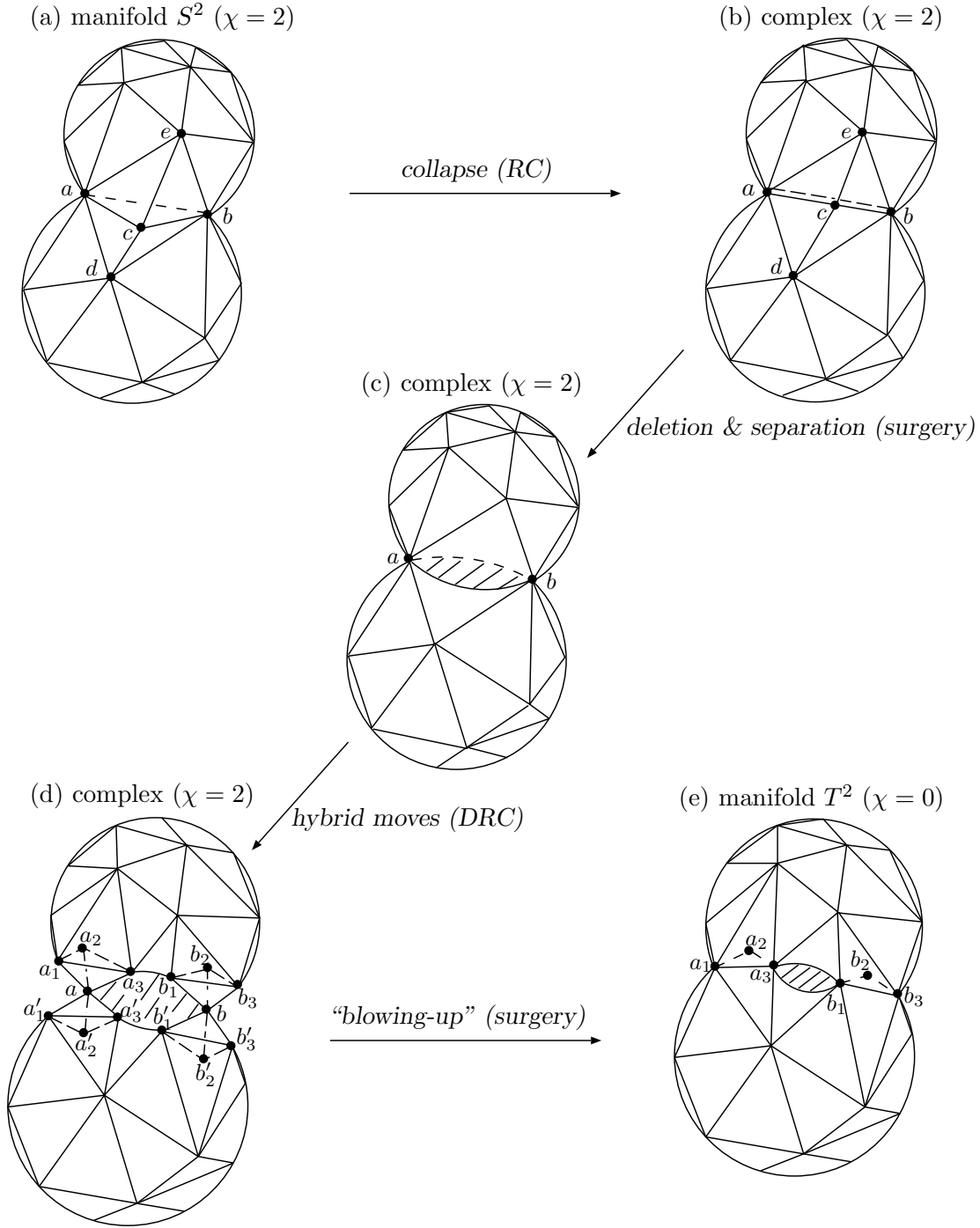}}
\put(1.0,18.3){(a) manifold $S^2$ $(\chi = 2)$}
\put(2.97,16.59){\small $e$} \put(1.47,15.72){\small $a$} 
\put(2.40,15.05){\small $c$} \put(3.91,15.3){\small $b$} 
\put(1.84,14.51){\small $d$}
\put(6.6,16.0){\sl collapse (RC)}
\put(11.4,18.3){(b) complex $(\chi = 2)$}
\put(13.25,16.59){\small $e$} \put(11.75,15.72){$a$}
\put(12.88,15.36){\small $c$} \put(14.19,15.3){\small $b$} 
\put(12.18,14.53){\small $d$}
\put(10.,12.1){\sl deletion \& separation (surgery)}
\put(6.0,12.77){(c) complex $(\chi = 2)$}
\put(6.39,10.16){\small $a$} \put(8.8,9.77){\small $b$}
\put(4.95,6.2){\sl hybrid moves (DRC)}
\put(1.0,6.55){(d) complex $(\chi = 2)$}
\put(1.14,3.93){\small $a_1$} \put(1.75,4.44){\small $a_2$}
\put(2.4,3.73){\small $a_3$} \put(1.53,3.62){\small $a$} 
\put(0.90,3.33){\small $a_1^{\prime}$} \put(1.68,2.8){\small $a_2^{\prime}$}
\put(2.36,3.31){\small $a_3^{\prime}$}
\put(2.94,3.65){\small $b_1$} \put(3.42,4.14){\small $b_2$} 
\put(4.14,3.49){\small $b_3$} \put(3.82,3.24){\small $b$}
\put(2.86,3.20){\small $b_1^{\prime}$} \put(3.5,2.35){\small $b_2^{\prime}$}
\put(4.12,2.89){\small $b_3^{\prime}$}
\put(5.5,3.2){\sl ``blowing-up'' (surgery)}
\put(10.8,6.5){(e) manifold $T^2$ $(\chi = 0)$}
\put(10.96,4.03){\small $a_1$} \put(11.8,4.38){\small $a_2$}
\put(11.87,3.87){\small $a_3$}
\put(13.03,3.4){\small $b_1$} \put(13.39,4.07){\small $b_2$}
\put(14.2,3.52){\small $b_3$}
\end{picture}
\end{center}
\vspace*{-0.9cm}
\caption
{{\sl 
Example of a topology-changing process via degenerate complexes in extended 
DRC {\rm (\ref{eq:Z_DRC_ext})}. {\rm (a)} A simplicial manifold $S^2$ of 
$\chi=2$ has a `waist' triangle $abc$. {\rm (b)} If three links of $abc$ 
satisfy $l_{ab} = l_{ac} + l_{bc}$, then $abc$ collapses to a (singular) link. 
The configuration is now {\it not\/} a manifold 
{\it but} a degenerate complex. {\rm (c)} The four 2-simplices sharing the 
0-simplex $c$ are deleted. But we leave the two 1-simplices $ac$ and $bc$ and 
combine them into another 1-simplex $ab$ (solid curve), and then delete
$c$. The dashed curve is the old $ab$. The shaded region between the two $ab$ 
becomes empty. {\rm (d)} By acting several hybrid $(p,\ts q)$ moves 
around $a$ and $b$, the two vertices will have coordination number three. The 
complex still has $\chi=2$. {\rm (e)} We delete six 2-simplices around $a$ and 
further six 2-simplices around $b$. Then, the 0-simplices $a_i$ and
$a_i^{\prime} \es (i=1,2,3)$ are identified and, similarly, $b_i$ and
$b_i^{\prime} \es (i=1,2,3)$ are done. The complex has changed to a simplicial 
manifold $T^2$ of $\chi=0$ by the ``blowing-up'' surgery. 
}}
\label{fig:tplgyChngRC}
\end{figure}

  Now we discuss a simple example in Fig.~\ref{fig:tplgyChngRC} to illustrate 
a process of changing continuously from the topology of one simplicial 
manifold to the topology of another.
Consider a simplicial manifold $S^2$ with a `waist' triangle $abc$ 
(Fig.~\ref{fig:tplgyChngRC}~(a)). The three links $ab, ac$ and $bc$ of the
`waist' has lengths $l_{ab}, l_{ac}$ and $l_{bc}$ respectively, satisfying a
triangle inequality $l_{ab} < l_{ac} + l_{bc}$. The manifold has Euler number 
$\chi=2$, where $\chi$ is defined on the two-dimensional lattice as
\begin{eqnarray*}
\chi = N_0 - N_1 + N_2 \es.
\end{eqnarray*}
Here $N_k$ stands for the number of $k$-simplices $(k=0,1,2)$. Once an equality
$l_{ab} = l_{ac} + l_{bc}$ holds by varying the link-lengths, the triangle 
$abc$ will collapse to a singular link along which the manifold condition does 
not hold any longer (Fig.~\ref{fig:tplgyChngRC}~(b)). In general, if some of 
the (sub)simplices collapse to lower dimensions, a simplicial complex will not 
remain a manifold but become a degenerate configuration; this is the case with 
the simplicial complex shown in Fig.~\ref{fig:tplgyChngRC}~(b). 

  Subsequently, we delete four 2-simplices $adc, bdc, bec$ and $eac$ around 
$c$, while combining the two links $ac$ and $cb$ into a new link $ab$. Then, 
$c$ is deleted. Moreover, a separation of the new link $ab$ (dashed) from the 
old $ab$ (solid) will lead to other degenerate complex shown in 
Fig.~\ref{fig:tplgyChngRC}~(c). The shaded part is an empty region on which no 
(sub)simplex resides. Although the simplicial complex is no longer a manifold, 
the Regge action (\ref{eq:S_Regge}) remains well-defined and finite even on 
such degenerate configurations shown in Figs.~\ref{fig:tplgyChngRC}~(b) and 
(c). The process from (b) to (c) is beyond the link-integrations (of RC) and 
the hybrid $(p,\ts q)$ moves (of DRC); in general, such a process 
should be called a {\em surgery\/} acting on the simplicial complex.
%
%

  Next step is to apply hybrid $(p,\ts q)$ moves several times to the simplices
around $a$ and $b$. As a result, we can obtain a simplicial complex as shown in
Fig.~\ref{fig:tplgyChngRC}~(d), in which coordination numbers of $a$ and $b$ 
are three. However, the obtained complex still has apparently Euler number 
$\chi=2$, though it is {\em not\/} a manifold of definite dimension and 
topology.

  Finally, we ``blow up'' the degenerate simplicial complex to obtain a new 
simplicial manifold. We delete six 2-simplices around $a$, that is, 
$aa_1a_2,\ts aa_2a_3,\ts aa_3a_1,\ts aa_1^{\prime}a_2^{\prime},\ts 
aa_2^{\prime}a_3^{\prime}$ and $aa_3^{\prime}a_1^{\prime}$ in 
Fig.~\ref{fig:tplgyChngRC}~(d). Then $a$ is deleted, while we identify $a_i$ 
with $a_i^{\prime}\es (i=1,2,3)$ pairwise. Similarly, we delete six 
2-simplices around $b$, and identify $b_i$ with $b_i^{\prime} \es (i=1,2,3)$. 
As a consequence of the ``blowing-up'' surgery, the degenerate complex of 
Fig.~\ref{fig:tplgyChngRC}~(d) has changed to a simplicial manifold $T^2$ of 
Fig.~\ref{fig:tplgyChngRC}~(e). 

  In this way, one can pass continuously from one metric topology to another 
with the Regge action remaining finite. Indeed, one could deform the topology 
of the simplicial manifold only by the action of such local moves within the 
framework of extended DRC (\ref{eq:Z_DRC_ext}). However, we meet with a 
difficulty in finding a minimal set of local moves that are enough to generate
step by step all the (degenerate) complexes of various topologies. 
This is due to the surgery operations, such as the process from
Fig.~\ref{fig:tplgyChngRC}~(b) to (c) or the process from 
Fig.~\ref{fig:tplgyChngRC}~(d) to (e). Therefore, it is a challenging task
for us to study numerically the extended model (\ref{eq:Z_DRC_ext}).

%
\vspace*{0.4cm}
\section{Lattice diffeomorphisms in DRC}\label{sect:genCovLat}
\spc Given a Riemannian manifold $(M,\ts g)$, there is arbitrariness in 
choosing coordinate-systems that cover the manifold $M$. The arbitrariness is 
necessary to represent the principle of general covariance in general 
relativity. Similarly, in the case of lattice gravity, the freedom would be 
reflected as the existence of the infinite number of triangulations that 
correspond to the same Riemannian geometry; they should be transformed to each 
other by ``lattice diffeomorphisms'' defined on the simplicial manifold.
An important issue in regularized theories of quantum gravity is the nature of 
the symmetric properties. Though significant discussions on the issue have been
given, no systematic formulations of the symmetry have been established. 
Actually, in the FT approach diffeomorphism-invariance is approximately 
realized as the appearance of gauge zero modes only on (almost) flat 
backgrounds~\cite{HW2,HW3,RW}; by the method of lattice weak-field expansion, 
the zero modes and the corresponding eigenvectors identified with infinitesimal
local coordinate transformations in a continuum limit. But invariance 
properties on curved backgrounds are beyond the scope 
of the weak-field expansion. 

  In this section we will see that the Regge action (\ref{eq:S_Regge}) on 
arbitrary curved backgrounds is exactly invariant under certain hybrid 
$(p,\ts q)$ moves of (extended) DRC and further that such moves are interpreted
as the lattice diffeomorphisms in a natural way.
%
\subsection{Invariance $(p,\ts q)$ moves on the simplicial complex}
\begin{figure}[h]
\vspace*{0.0cm}
\begin{center}
 \setlength{\unitlength}{1cm}
 \begin{picture}(15.0,4.8)(0,0)
 \put(0.5,0.0){\includegraphics[width=14cm]{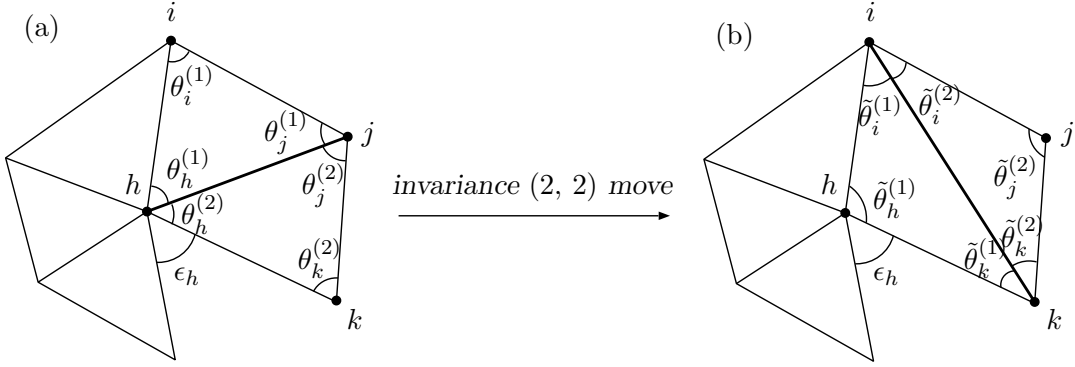}}
 \put(0.8,4.4){(a)}
 \put(2.8,1.15){$\epsilon_h$}
 \put(2.7,4.6){$i$} \put(2.15,2.3){$h$} 
 \put(5.3,3.0){$j$} \put(5.1,0.5){$k$}
 \put(2.75,3.6){$\theta_i^{(1)}$} \put(2.7,2.5){$\theta_h^{(1)}$}
 \put(4.0,3.0){$\theta_j^{(1)}$} \put(2.9,1.9){$\theta_h^{(2)}$}
 \put(4.5,2.3){$\theta_j^{(2)}$} \put(4.45,1.3){$\theta_k^{(2)}$}
 \put(5.7,2.3){\sl invariance $(2,\ts 2)$ move}
 \put(10.0,4.3){(b)}
 \put(12.1,1.15){$\epsilon_h$}
 \put(12.0,4.6){$i$} \put(11.4,2.3){$h$} 
 \put(14.6,3.0){$j$} \put(14.4,0.5){$k$}
 \put(11.9,3.2){$\tilde{\theta}_i^{(1)}$}
 \put(12.7,3.35){$\tilde{\theta}_i^{(2)}$}
 \put(13.7,2.4){$\tilde{\theta}_j^{(2)}$}
 \put(12.1,2.1){$\tilde{\theta}_h^{(1)}$}
 \put(13.3,1.3){$\tilde{\theta}_k^{(1)}$}
 \put(13.8,1.60){$\tilde{\theta}_k^{(2)}$}
 \end{picture}
\end{center}
\vspace*{-0.8cm}
\caption
{{\sl 
Example of the invariance (2, 2) move which keeps the simplicial geometry and 
the Regge action (\ref{eq:S_Regge}) invariant. {\rm (a)} Some triangles sharing
the hinge $h$ are put on a flat $\mbox{\boldmath $R$}^2$ plane in order
to visualize the deficit angle $\epsilon_h$. {\rm (b)} The link $hj$ is flipped
to other link $ik$ in such a way that the length of the link $ik$ is the 
Euclidean distance between the vertices $i$ and $k$.
}}
\label{fig:inv_2_2move}
\vspace{0.5cm}
\end{figure}

\spc As an illustrative example, we first construct a hybrid $(2,\ts 2)$ 
move under which the Regge action (\ref{eq:S_Regge}) remains exactly invariant.
We zoom up some 2-simplices around a hinge (vertex) $h$ on a triangulated 
surface, as shown in Fig.~\ref{fig:inv_2_2move}.
%
We put them on a flat $\mbox{\boldmath $R$}^2$ plane to make explicitly visible
the deficit angle $\epsilon_h$ around $h$, as depicted on 
Fig.~\ref{fig:inv_2_2move}~(a). A hybrid $(2,\ts 2)$ move makes the link $hj$ 
flip to other link $ik$ in such a way that the vertices $i$ and $k$ are 
connected by a straight line on $\mbox{\boldmath $R$}^2$, 
as shown in Fig.~\ref{fig:inv_2_2move}~(b). 
Evidently, the deficit angles $\epsilon_h,\ts \epsilon_i,\ts \epsilon_j,\ts
\epsilon_k$ around $h,\ts i,\ts j,\ts k$ are invariant,
because the following relations hold:
\begin{eqnarray}
\theta_h^{(1)} +  \theta_h^{(2)} = \tilde{\theta}_h^{(1)} \ts, \hspace*{1cm}
\theta_i^{(1)} = \tilde{\theta}_i^{(1)} + \tilde{\theta}_i^{(2)} \ts, 
\nonumber \\
\theta_j^{(1)} +  \theta_j^{(2)} = \tilde{\theta}_j^{(2)} \ts, \hspace*{1cm}
\theta_k^{(1)} = \tilde{\theta}_k^{(1)} + \tilde{\theta}_k^{(2)} \ts.\ts
\label{eq:dihed_2_2}
\end{eqnarray}
where $\theta_v^{(m)}$ and $\tilde{\theta}_v^{(m)} \es (v = h,i,j,k;\es m=1,2)$
are shown in Fig.~\ref{fig:inv_2_2move}~(a) and (b).
The sum of areas of the triangles also is invariant:
\begin{eqnarray}
V_{hij} + V_{hkj} = V_{hik} + V_{ijk} \es.
\label{eq:vol_2_2}
\end{eqnarray}
Here the volumes $V_{\sigma} \es (\sigma = hij,\ts hkj,\ts hik,\ts ijk)$ denote
areas of the 2-simplices considered. These simple relations 
(\ref{eq:dihed_2_2}) and (\ref{eq:vol_2_2}) ensure that the local geometry and 
the Regge action (\ref{eq:S_Regge}) are exactly invariant under the move of
Fig.~\ref{fig:inv_2_2move}. Hence, we call this hybrid $(2,\ts 2)$ move an 
{\em invariance $(2,\ts 2)$ move}. From a geometric viewpoint, the invariance
$(2,\ts 2)$ move is a simplicial representation of a local coordinate 
transformation, because the 2-simplices $hij,\ts hjk,\ts ihk$ and $ijk$ 
themselves can be regarded as local coordinates and the move changes one local 
coordinate system spanned by $hij$ and $hjk$ to another spanned by $ihk$ and 
$ijk$. Furthermore, it keeps exactly invariant the simplicial geometry in the 
simple way. Therefore, we can naturally interpret the invariance $(2,\ts 2)$ 
move as a lattice diffeomorphism in the sense of 
``invariance of the geometry''.
%
%
\begin{figure}[h]
\vspace*{0.7cm}
\begin{center}
 \setlength{\unitlength}{1cm}
 \begin{picture}(16.0,5.0)(0,0)
 \put(0.5,0.4){\includegraphics[width=15cm]{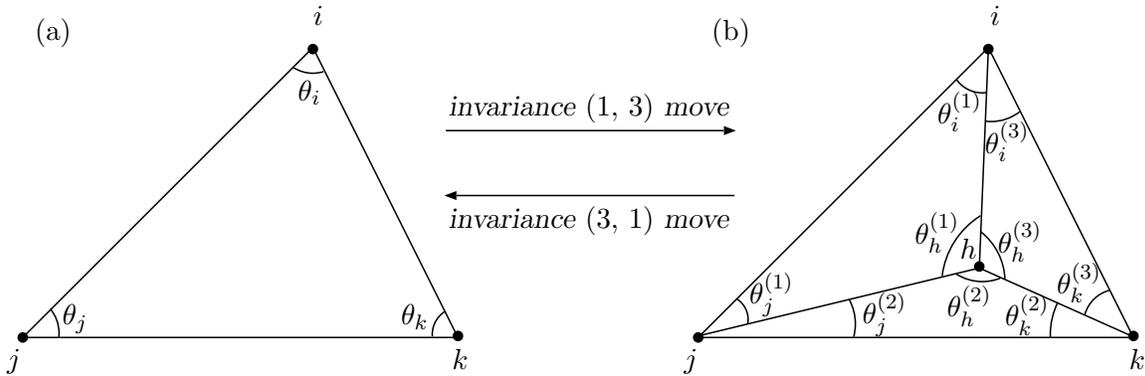}}
 \put(0.8,4.5){(a)}
 \put(4.5,4.7){$i$} \put(0.43,0.14){$j$} \put(6.35,0.13){$k$}
 \put(4.3,3.7){$\theta_i$} \put(1.17,0.72){$\theta_j$} 
 \put(5.67,0.72){$\theta_k$}
 \put(6.3,3.5){\sl invariance $(1,\ts 3)$ move}
 \put(6.3,2.0){\sl invariance $(3,\ts 1)$ move}
 \put(9.8,4.5){(b)}
 \put(13.5,4.7){$i$} \put(9.43,0.14){$j$} \put(15.35,0.13){$k$}
 \put(12.8,3.4){$\theta_i^{(1)}$} \put(13.43,3.0){$\theta_i^{(3)}$}
 \put(10.3,1.0){$\theta_j^{(1)}$} \put(12.5,1.8){$\theta_h^{(1)}$}
 \put(13.1,1.6){$h$}
 \put(11.8,0.73){$\theta_j^{(2)}$} \put(12.93,0.82){$\theta_h^{(2)}$}
 \put(13.6,1.68){$\theta_h^{(3)}$} \put(14.39,1.13){$\theta_k^{(3)}$}
 \put(13.7,0.7){$\theta_k^{(2)}$}
 \end{picture}
\end{center}
\vspace*{-0.8cm}
\caption
{{\sl
Example of the invariance $(1,\ts 3)$ move, which keeps invariant the 
simplicial geometry and the Regge action (\ref{eq:S_Regge}). {\rm (a)} A 
triangle $ijk$ is put on a flat $\mbox{\boldmath $R$}^2$ plane. {\rm (b)} The 
triangle is subdivided into three triangles $hij,\ts hjk$ and $hki$ 
in such a way that the length of the link $hi$ is the Euclidean distance 
(straight line) between $h$ and $i$ and, in the similar way, the lengths of 
$hj$ and $hk$ are chosen to be the Euclidean distances.
}}
\label{fig:inv_1_3move}
\vspace{0.5cm}
\end{figure}

  Similarly, we can define other invariance moves as shown in
Fig.~\ref{fig:inv_1_3move}. 
We first pick up a 2-simplex $ijk$ shown in Fig.~\ref{fig:inv_1_3move}~(a), and
then subdivide it into three simplices $hij,\ts hjk$ and $hki$ in 
Fig.~\ref{fig:inv_1_3move}~(b) by putting a new vertex $h$ inside $ijk$ and 
connecting it with $i,\ts j,\ts k$ in such a way that the flatness property 
inside $ijk$ is preserved. Under the action of this move, the invariance of the
deficit angles $\epsilon_i,\ts \epsilon_j,\ts \epsilon_k$ around
$i,\ts j,\ts k$ is guaranteed simply by the following relations:
\begin{eqnarray}
\theta_i =  \theta_i^{(1)} + \theta_i^{(3)} \ts, \es\es
\theta_j =  \theta_j^{(1)} + \theta_j^{(2)} \ts, \es\es
\theta_k =  \theta_k^{(2)} + \theta_k^{(3)} \ts.
\label{eq:dihed_1_3}
\end{eqnarray}
Here, the dihedral angles $\theta_v$ and $\theta_v^{(m)} \es (v = i,j,k;\ts 
m=1,2,3)$ are shown in Fig.~\ref{fig:inv_1_3move}~(a) and (b). The deficit 
angle $\epsilon_h$ around the inserted vertex $h$ is exactly zero, because 
the Euclidean flat geometry inside the triangle $ijk$ is preserved exactly,
namely, the following equality holds:
\begin{eqnarray}
2\pi =  \theta_h^{(1)} +  \theta_h^{(2)} + \theta_h^{(3)} \ts. 
\label{eq:dihed_1_3_flat}
\end{eqnarray}
The total volume also is invariant under the move:
\begin{eqnarray}
V_{ijk} = V_{hij} + V_{hjk} + V_{hki} \es,
\label{eq:vol_1_3}
\end{eqnarray}
where the volumes $V_{\sigma}$ denote areas of the 2-simplices $\sigma = ijk,
\ts hij,\ts hjk$ and $hki$ in Fig.~\ref{fig:inv_1_3move}. We call this hybrid 
$(1,\ts 3)$ move an {\em invariance $(1,\ts 3)$ move}, and its inverse an 
{\em invariance $(3,\ts 1)$ move}. These relations (\ref{eq:dihed_1_3}), 
(\ref{eq:dihed_1_3_flat}) and (\ref{eq:vol_1_3}) guarantees the exact 
invariance of the local geometry under the moves. These moves can be
interpreted as the lattice diffeomorphisms in the same way as the invariance 
$(2,\ts 2)$ move could be.

  It is straightforward to generalize the two-dimensional invariance 
$(p,\ts q)$ moves to higher-dimensional ones, under which both the simplicial
geometry and the Regge action are kept exactly invariant in the similar way. 
For example, we consider the three-dimensional case. Let us remember 
Fig.~\ref{fig:ex_kl_moves_3}, where the 3-simplex $v_0v_1v_2v_3$ is divided 
into the four 3-simplices by the hybrid $(1,\ts 4)$ move. If one adjusts the 
lengths $l_0,\ts l_1,\ts l_2$ and $l_3$ so that the geometry inside 
$v_0v_1v_2v_3$ remains flat, the deficit angles around the six links of 
$v_0v_1v_2v_3$ are invariant and, furthermore, deficit angles around the four 
inserted links are also zero (flat) in Fig.~\ref{fig:ex_kl_moves_3}~(b). The 
sum of volumes of the four new 3-simplices is exactly equal to that of 
$v_0v_1v_2v_3$. This action is an invariance $(1,\ts 4)$ move that keeps the 
Regge action (\ref{eq:S_Regge}) invariant in three dimensions. Similarly, we
can construct an invariance $(2,\ts 3)$ and an invariance $(3,\ts 2)$ moves.

  In addition to the symmetry described above, the permutation group,
$S_{N_0}$, of $N_0$ vertices gives another invariance, corresponding to the
re-labeling of vertices on the lattice. Therefore, we conclude that 
{\em the lattice diffeomorphisms in the sense of invariance of the geometry
are generated by both the invariance $(p,\ts q)$ moves and the permutation
group $S_{N_0}$ in DRC.$\ts$} 
Symbolically, we denote the symmetry of DRC as follows:
\begin{eqnarray}
\Bigl\{ \mbox{lattice diffeomorphisms} \Bigr\} = 
\Bigl\{ \mbox{moves generated by invariance $(p,\ts q)$ moves} \Bigr\} 
\ts {\Large \ltimes} \bigoplus_{N_0=1}^{\infty}S_{N_0} \es.
\label{eq:LatDiff}
\end{eqnarray}
%


  What does this symmetry imply to lattice quantum gravity?  One might naively 
expect that one can simply integrate over the gauge transformations (the 
lattice diffeomorphisms) without any gauge-fixing procedure just as one could
in lattice gauge theories. However, this is {\em not\/} the case with 
gravitation~\cite{RW1}; the crucial difference between Einstein's gravity and 
gauge theories is that the symmetry group is {\em non-compact\/} in theories of
(lattice) gravity. Thereby, one must factor out the {\em infinite\/} 
volume of the diffeomorphism group to make sense of the path integral for the
gravitational field. Thus, it is plausible that the infinite volume of the 
lattice diffeomorphisms would prevent the entropy bound (\ref{eq:W_bound}) 
from holding. Generally speaking, it is very difficult to prove (or deny) the 
inequality (\ref{eq:W_bound}) by analytic methods, because the entropy 
(\ref{def:entropy_DRC}) depends on the triangle inequalities which are too 
complicated to estimate analytically. Moreover, it will also depend on the 
measure chosen for the link-integration. Hitherto, we deform the problem to 
another one: what measure enables the exponential bound (\ref{eq:W_bound}) to 
hold?  In section~\ref{sect:numSimln}, we will investigate numerically the 
(un)boundedness of the entropy (\ref{def:entropy_DRC}) using a few types of 
measures.
%
\subsection{Black hole entropy in DRC}
\spc Before turning to numerical studies, let us now apply the lattice symmetry
to the black hole entropy problem.
Among various attempts to solve the problem,
Carlip advocated that by virtue of the horizon structure, the ``would-be gauge 
freedom'' of general covariance supplies physical states of the Kerr black 
holes in arbitrary dimensions~\cite{Carlip}. In spite of some shortcomings in 
his analysis~\cite{CrlpSCM}, his idea has attracted much attention to the 
relation between the entropy and the horizontal geometry of black holes.
Indeed, it is argued in ref.~\cite{HSS} that the smooth diffeomorphism on the 
event horizon can be regarded as a nontrivial asymptotic isometry of the
Schwarzschild black hole. In what follows, we discuss a simple method of 
deriving the black hole entropy from a lattice-theoretic viewpoint; the 
derivation is based on the lattice diffeomorphism (\ref{eq:LatDiff}) in DRC.

\begin{figure}
\vspace*{0.0cm}
\begin{center}
 \setlength{\unitlength}{1cm}
 \begin{picture}(13.0,7.0)(0,0)
 \put(0.5,0.0){\includegraphics[width=12cm]{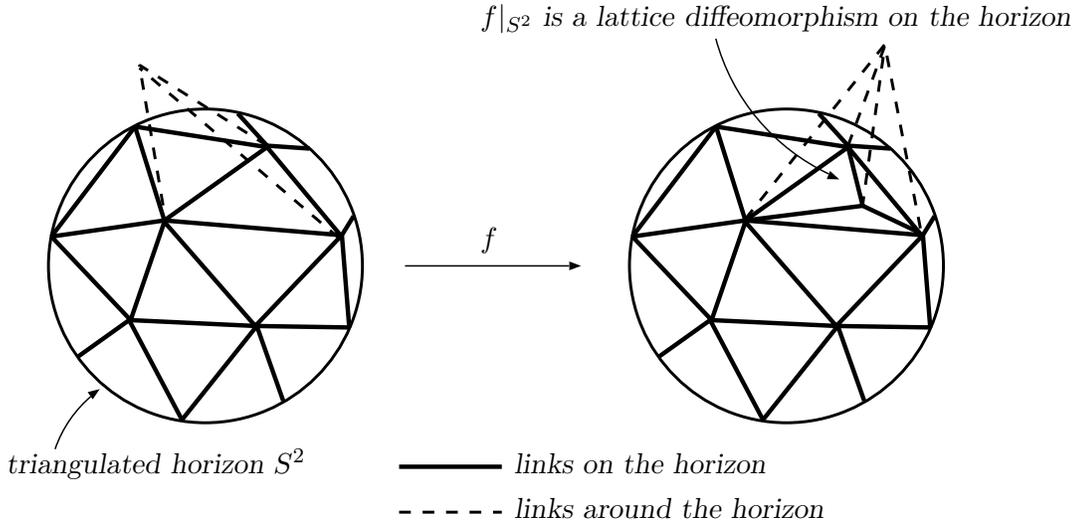}}
 \put(6.3,6.55){\sl $f|_{S^2}$ is a lattice diffeomorphism on the horizon}
 \put(0.0,0.6){\sl triangulated horizon $S^2$} \put(6.3,3.6){$f$}
 \put(6.75,0.6){\sl links on the horizon}
 \put(6.75,0.0){\sl links around the horizon}
 \end{picture}
\end{center}
\vspace*{-0.5cm}
\caption
{{\sl 
Example of a combination of the hybrid moves $f \in H_{HM}$ defined by
eq.~(\ref{def:HMevhrznDiff}). On the triangulated horizon $S^2$, 
$f |_{S^2}$ is required to be a two-dimensional lattice diffeomorphism.
}}
\label{fig:hybridMoveOnHorizen}
\vspace*{0.5cm}
\end{figure}

  Suppose that there exists a simplicial lattice that corresponds to a 
four-dimensional black hole with a triangulated horizon $S^2$ of area $A$.
We define $G_{\rm HM}$ as a set of all the moves generated by the hybrid 
$(p,\ts q)$ moves acting on the triangulated black hole. 
We define a subset $H_{\rm HM} \subset G_{\rm HM}$ as
\begin{eqnarray}
H_{\rm HM} 
&\equiv& \Bigl\{ f \in G_{\rm HM} \es ; \es f \notin
\bigl\{ \mbox{lattice diffeomorphisms} \bigr\}, \es 
f|_{S^2} \ne \mbox{identity},\es \mbox{and} \hspace*{2.0cm} \es \nonumber \\
&      & \hspace*{1.5cm} \es f|_{S^2} \in \bigl\{ \mbox{lattice 
diffeomorphisms on the triangulated horizon $S^2$ } \bigr\} \Bigr\} \ts.
\label{def:HMevhrznDiff}
\end{eqnarray}
An example of an element $f \in H_{\rm HM}$ is depicted on 
Fig.~\ref{fig:hybridMoveOnHorizen}. Physically, each element $f \in H_{\rm HM}$
means such a quantum fluctuation that is a nontrivial combination of the hybrid
$(p,\ts q)$ moves acting on the triangulated horizon; its restriction,
$f|_{S^2}$, on the horizon is required to be a two-dimensional lattice 
diffeomorphism to keep invariant the horizontal geometry.

  Assuming that there exist $n_{\rm P}$ elements of $H_{\rm HM}$ per the Planck
area $\ell_{\rm P}^2 = G \hbar / c^3$, we can easily count 
the total number of such fluctuations, $N_{fluc}$, around the horizon 
in a combinatorial way:
\begin{eqnarray*}
N_{fluc} \sim\ts \left(n_{\rm P}\right)^{A /\ell_{\rm P}^2} \es.
\end{eqnarray*}
This number $N_{fluc}$ corresponds to the total number of quantum fluctuations
that keep invariant the simplicial geometry on the horizon. Hence,
we can define the entropy, ${\cal S}_{\rm BH}$, of the black hole as
\begin{eqnarray}
{\cal S}_{\rm BH} \equiv k_{\rm B} \log N_{fluc} \sim\ts k_{\rm B} 
\log n_{\rm P} \cdot \frac{\ds A}{\ds \ell_{\rm P}^2} \es.
\label{eq:entBHfinite}
\end{eqnarray}

  Our idea of the derivation of the black hole entropy (\ref{eq:entBHfinite})
is based on the simple picture that any quantum states are associated with 
quantum fluctuations; the classical picture of the horizon should accompany 
many fluctuations around it at a quantum level, and the huge entropy of the 
black hole arises from the contribution of such fluctuations.

  Here, we encounter a difficult problem: If one takes a continuum limit, 
the fluctuation density $n_{\rm P}$ will diverge, resulting in an infinite 
value of the entropy ${\cal S}_{\rm BH}$ in eq.~(\ref{eq:entBHfinite}). 
%
%
  Such a divergent behavior of the black hole entropy often appears also in 
continuum approaches.
In particular, Susskind and Uglum~\cite{SusskindUglum} showed that the entropy 
per unit area of a free scalar field propagating in a fixed black hole 
background is quadratically divergent near the horizon and such quantum 
corrections to the black hole entropy are equivalent to quantum 
corrections to the gravitational coupling constant $G$, although the theory 
of interest is non-renormalizable. They concluded that the question on the 
finiteness of the entropy is inextricably intertwined with the renormalization 
of the coupling $G$ and, therefore, the finiteness property cannot be 
understood without the complete knowledge of the ultra-violet behavior 
of the theory.

  Taking the discussions in the continuum into consideration, we are led to
the following idea: If we regard the lattice model (\ref{eq:Z_DRC}) as an 
effective theory with a finite cutoff $l_{\rm min}^{-1}$, the fluctuation 
density $n_{\rm P}$ will remain finite in eq.~(\ref{eq:entBHfinite}). This 
finiteness enables us to avoid the divergence behavior and, furthermore, to 
normalize the Planck length $\ell_{\rm P}$ in eq.~(\ref{eq:entBHfinite}) 
so as to set 
\begin{eqnarray}
n_{\rm P} = \ts e^{1/4} \es (= 1.284 \ldots) \es.
\label{eq:nrmzFlcDnst}
\end{eqnarray}
As a consequence, we reach the following expression:
\begin{eqnarray}
{\cal S}_{\rm BH} \sim\ts \frac{\ds k_{\rm B}}{\ds 4} 
\frac{\ds A}{\ds \ell_{\rm P}^2} \es.
\label{eq:BHentropy}
\end{eqnarray}
Thus, eq.~(\ref{eq:BHentropy}) agrees with the Bekenstein-Hawking area 
law of the black hole entropy (\ref{eq:entropyBH}).

%
%
  How can we interpret the relations (\ref{eq:nrmzFlcDnst}) and 
(\ref{eq:BHentropy}) on physical ground? Interestingly, the notion of 
the {\em space-time uncertainty principle\/} has been advocated in the 
development of string theory and noncommutative field theory~\cite{yoneya}. 
According to the principle, space-time has an uncertain property in itself;
uncertainty of temporal coordinate ${\mit\Delta} T$ and that of spatial 
one ${\mit\Delta} X$ satisfy
\begin{eqnarray}
{\mit\Delta} T \cdot {\mit\Delta} X \ge \ell_{\rm s}^{2} \es,
\label{eq:spcTmUncertain}
\end{eqnarray}
where $\ell_{\rm s}$ is the so-called string scale of order $O(\ell_{\rm P})$. 
The lengths ${\mit\Delta} T$ and ${\mit\Delta} X$, called the extremal length,
are defined in a conformally invariant way~\cite{yoneya}. An interpretation of 
the relation (\ref{eq:spcTmUncertain}) is given in the context of 
noncommutative geometry~\cite{NCG}, in which the coordinates are promoted to 
operators that satisfy commutation relations similar to those of quantum
mechanics.
However, we take another interpretation of the uncertainty principle
(\ref{eq:spcTmUncertain}); space-time has the intrinsic minimal 
length\footnote{Actually, it is discussed in ref.~\cite{GKP} 
that a particular uncertainty relation leads to the minimal length for each
component of the space-time coordinates, namely, eq.~(\ref{eq:spcTmUPtrvl})
may be satisfied also in string theory.} $\ell_{\rm s}$ and
we identify it with the UV cutoff $l_{\rm min}$ of DRC:
\begin{eqnarray}
{\mit\Delta} T_{\mbox{\sp min}} = {\mit\Delta} X_{\mbox{\sp min}} 
= \ell_{\rm s} \sim l_{\rm min} \es.
\label{eq:spcTmUPtrvl}
\end{eqnarray}
As a result, the uncertainty principle can be satisfied without introducing
noncommutativity of space-time. According to this interpretation,
eq.~(\ref{eq:nrmzFlcDnst}) means that on the minimal area $l_{\rm min}^2 \sim
\ell_{\rm P}^2$ we observe about one quantum fluctuation of the (lattice) 
gravitational field.
\vspace*{0.6cm}
\section{Numerical study of 3D pure gravity}\label{sect:numSimln}
\spc In this section we turn to numerical calculations in order to study
non-perturbatively the ground state of DRC (\ref{eq:Z_DRC}) and the behavior 
of the entropy (\ref{def:entropy_DRC}). We will calculate some observables
(integrated curvature, average link-length, and surface area distribution
function characterizing a fractal structure) to study
three-dimensional pure gravity based on DRC.
%
%

  First, we rewrite the measure $d\mu_T[l]$ in the following form
convenient for our numerical study:
\begin{eqnarray}
\int d\mu_T[l] = \prod_{i=1}^{N_1} \int_{l_{\rm min}}^{l_{\rm max}} \nts\nts 
d l_i \ts \exp \Bigl( - \tilde{S}[l,\ts T] \Bigr) \ts \delta_T(\Delta) \es,
\label{eq:S_measure}
\end{eqnarray}
%
where $\delta_T(\Delta)$ stands for the constraints by the (higher-dimensional 
analogs of) triangle inequalities, and we denote the measure-induced part of
the action as $e^{- \tilde{S}[l,\ts T]}$. Two examples of the measure-induced
part $\tilde{S}[l,\ts T]$ will soon appear.
  From now on we set the minimum lattice spacing $l_{\rm min}$ to be 
one\footnote{If dimensional arguments are needed, the appropriate power of the 
lattice spacing (which has the dimension of length) can always be restored at 
the end of calculations by invoking dimensional arguments. }.

  In the DT approach the number of $d$-simplices, $N_d$, fluctuates largely 
in higher dimensions $(d = 3,\ts 4)$, making statistics of numerical data 
fairly worse. Similarly, such fluctuations occur also in our hybrid model. In 
order to obtain better statistics, we add an extra term $(N_d - V)^2$ to the 
Regge action (\ref{eq:S_Regge}); it controls the volume fluctuations around the
designed size $V$. Hence, the total action, $S[l,\ts T]$, on the
lattice is of the form:
\begin{eqnarray}
S[l,\ts T] = S_{\mbox{\sp Regge}}[l,\ts T]
+ \tilde{S}[l,\ts T]
+ \gamma (N_d - V)^2 \es.
\label{eq:Regge_act_eff}
\end{eqnarray}
Here $\gamma$ is a small constant. As a result, the partition function of
DRC becomes
\begin{eqnarray}
Z_{\mbox{\sp DRC}} = \sum_{T{\rm :triangulations}} \nts
\int \prod_i d l_i \ts
\exp \Bigl( -S[l,\ts T] \Bigr) \ts \delta_T(\Delta) \es,
\label{eq:Z_DRC_eff}
\end{eqnarray}
where the sum $\sum_{T{\rm :triangulations}}$ admits only simplicial manifolds
satisfying the manifold condition. Numerically, the sum of lattice 
configurations appearing in eq.~(\ref{eq:Z_DRC_eff}) is carried out through 
the hybrid $(p,\ts q)$ moves, as explained in section~\ref{sect:hybridMoves} 
(see Fig.~\ref{fig:ex_kl_moves_3}).

  From now on we focus on three-dimensional pure gravity. The topology of the 
simplicial lattice is fixed to three-sphere $S^3$ and we never consider the
topology-changing process here; this is due to the algorithmic difficulties in 
simulating manifolds of various topologies, which remains to be solved.
%
%
\subsection{Phase structure of 3D pure gravity}\label{sect:obsrvblsMC}
\spc In general, local operators cannot be gauge-invariant observables in 
quantum gravity. But some global quantities are gauge-invariant, for example, 
the average scalar curvature per volume:
\begin{eqnarray}
\langle R \rangle \equiv
\Biggl\langle \frac{\sum_{i} l_i \epsilon_i}
{\es \sum_{\sigma} V_{\sigma} \es} \Biggr\rangle  \es.
\label{def:ave_R}
\end{eqnarray}
Here $\epsilon_i$ is the deficit angle around the $i$-th link (hinge) $l_i$ 
and $V_{\sigma}$ being the volume of a 3-simplex $\sigma$. This quantity 
$\langle R \rangle$ plays the role of the order parameter.
The average length of all the links is also a good observable:
\begin{eqnarray}
\langle l \rangle \equiv
\Biggl\langle
\frac{\es \sum_{i} l_i \es}
{\es N_1 \es} \Biggr\rangle \es.
\label{def:ave_length}
\end{eqnarray}
These observables (\ref{def:ave_R}) and (\ref{def:ave_length}) can be 
calculated by the standard Monte-Carlo technique. In what follows, we perform 
numerical simulations using two types of measures, that is, the uniform measure
$\prod_i dl_i$ and the scale-invariant measure $\prod_i dl_i / l_i$. The number
of 3-simplices $N_3$ is almost fixed at around $V = 5000$, and each 
length $l_i$ is integrated over an interval $1 < l_i < 10$.
The coefficient of the extra term 
is set at a value $\gamma = 1.0 \times 10^{-5}$. 
%
\subsubsection{Calculation using the uniform measure}
\spc First, we use the uniform measure in simulating DRC (\ref{eq:Z_DRC}):
\begin{equation}
d\mu_T[l] = \prod_{i=1}^{N_1} dl_i 
\ts \delta_T(\Delta) \es.
\label{def:UniformMeasure}
\end{equation}
In this case, the lattice action $S[l,\ts T]$ is of the following simple form:
\begin{eqnarray}
S[l,\ts T] = S_{\mbox{\sp Regge}}[l,\ts T]
+ \gamma (N_3 - V)^2 \es.
\label{eq:Uniform_act_eff}
\end{eqnarray}
The partition function becomes
\begin{eqnarray}
Z_{\mbox{\sp DRC}}^{uniform} 
= \sum_{T} \int_1^{10} \prod_{i=1}^{N_1}  dl_i \ts \exp 
\Bigl( - S[l,\ts T] \Bigr) \ts \delta_T(\Delta) \es.
\label{eq:Z_DRC_uniform}
\end{eqnarray}
%
However, the measure (\ref{def:UniformMeasure}) causes a severe problem that 
the hybrid system (\ref{eq:Z_DRC_uniform}) will not be well-defined 
statistical-mechanically. This is due to the reason that the entropy bound 
(\ref{eq:W_bound}) for the configuration entropy (\ref{def:entropy_DRC}) will 
not hold in this case. Now we must discuss this point in more detail. 
%
%
\begin{figure}[h]
\vspace*{0.0cm}
\begin{center}
\includegraphics[height=8.0cm]{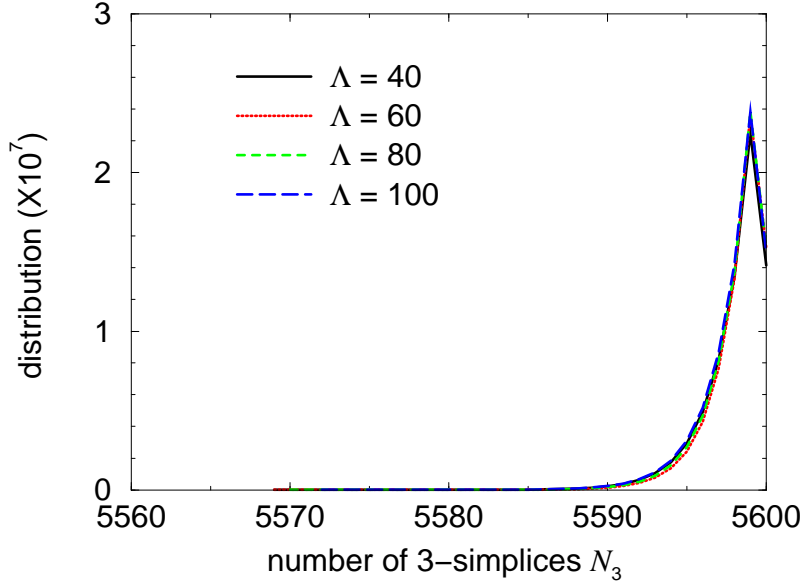}
\end{center}
\vspace*{-0.8cm}
\caption
{{\sl
The $N_3$ distribution using the uniform measure $\prod_i d l_i$. The 
coupling constant $G$ is set to be $\infty$. Different curves correspond to 
different values of the cosmological constant ranging from ${\mit\Lambda} = 40$
to ${\mit\Lambda} = 100$. The maximum value of the horizontal axis
is the upper cutoff of $N_3$.}}
\label{fig:UniMeasureCal}
\vspace*{0.5cm}
\end{figure}

  Actually, we tried carrying out numerical calculations using the measure 
(\ref{def:UniformMeasure}). In our numerical data with $G = \infty$ (strong 
coupling limit), however, the number of 3-simplices $N_3$ exhibits a clear 
tendency to diverge, even though we set large values of the cosmological 
constant ${\mit\Lambda}$, as shown in Fig.~\ref{fig:UniMeasureCal}.
Though the value of the cosmological constant varies from 40 to 100 in
Fig.~\ref{fig:UniMeasureCal}, the $N_3$ distribution is squeezed to the upper 
cutoff $N_3 = 5600$. Even if we choose larger values of the cutoff, the 
divergence behavior still appears in the same way. Hence, we cannot control 
the system (\ref{eq:Z_DRC_uniform}) at all. From a viewpoint of statistical 
mechanics, one can interpret such a pathological behavior as follows.
  If the entropy bound (\ref{eq:W_bound}) does not hold, the configuration
entropy $W(N_3)$ (given by eq.~(\ref{def:entropy_DRC})) will dominate the 
system in the strong-coupling phase (high-temperature phase). In other words, 
lattice configurations with large $N_3$ will frequently appear owing to the
large values of the entropy $W(N_3)$, leading to the divergence of the
lattice size. Actually, this is the case with our 
calculation using the uniform measure in Fig.~\ref{fig:UniMeasureCal}. 
%
The large volume of the lattice diffeomorphisms will make 
large (presumably infinite) the entropy of such pathological configurations.
In that sense the data of Fig.~\ref{fig:UniMeasureCal} indicates a piece of
numerical evidence for the existence of the lattice symmetry. 

   At any rate, our calculation using the uniform measure cannot proceed 
further. Instead of the pathological measure, we should next try to use other 
measure that may break the diffeomorphism invariance at the quantum level.
%
%
\subsubsection{Calculation using the scale-invariant measure}
\spc Next, we simulate the system (\ref{eq:Z_DRC_eff}) using 
the scale-invariant measure:
\begin{equation}
d\mu_T[l] = \prod_{i=1}^{N_1} 
\frac{\ds dl_i}{\ds l_i} \ts \delta_T(\Delta) \es.
\label{def:ConfMeasure}
\end{equation}
Obviously, this measure is invariant under the local rescalings $l_i \to c_i 
l_i$ where $c_i$ are constants. In this case, the partition function
is defined as
\begin{eqnarray}
Z_{\mbox{\sp DRC}}^{scale-inv} = 
\sum_{T} \int_1^{10} \prod_{i=1}^{N_1} dl_i
\ts \exp \Bigl( -S[l,\ts T] \Bigr) \ts \delta_T(\Delta) \es,
\label{eq:Z_DRC_cm}
\end{eqnarray}
where the action $S[l,\ts T]$ is of the form:
\begin{eqnarray}
S[l,\ts T] = S_{\mbox{\sp Regge}}[l,\ts T]
+ \sum_{i=1}^{N_1} \log l_i + \gamma (N_3 - V)^2 \es.
\label{eq:Conf_act_eff}
\end{eqnarray}
%

\begin{figure}[h]
\vspace*{0.4cm}
\begin{center}
\includegraphics[height=8.0cm]{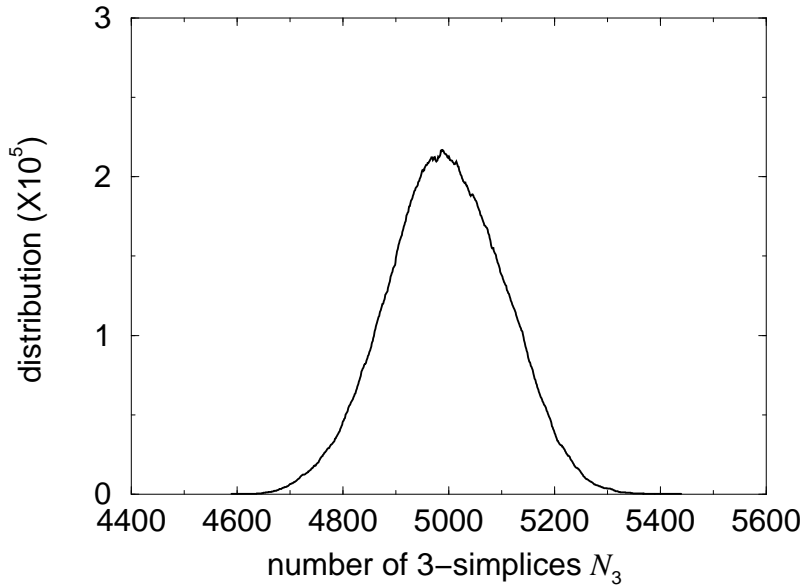}
\end{center}
\vspace*{-0.9cm}
\caption
{{\sl
$N_3$ distribution using the scale-invariant measure $\prod_i dl_i/l_i$.
}}
\label{fig:ConMeasureVol}
\vspace{0.1cm}
\end{figure}

  First of all, we measure the $N_3$ distribution, as shown in 
Fig.~\ref{fig:ConMeasureVol}. One observes in the figure that $N_3$ distributes
smoothly around the central value $V = 5000$, in contrast to the 
uniform-measure case shown in Fig.~\ref{fig:UniMeasureCal}. According to the 
numerical data of Fig.~\ref{fig:ConMeasureVol}, one can naively expect that the
exponential bound (\ref{eq:W_bound}) will hold under this measure.
\begin{figure}
\vspace{-0.0cm}
\begin{center}
\includegraphics[height=8cm]{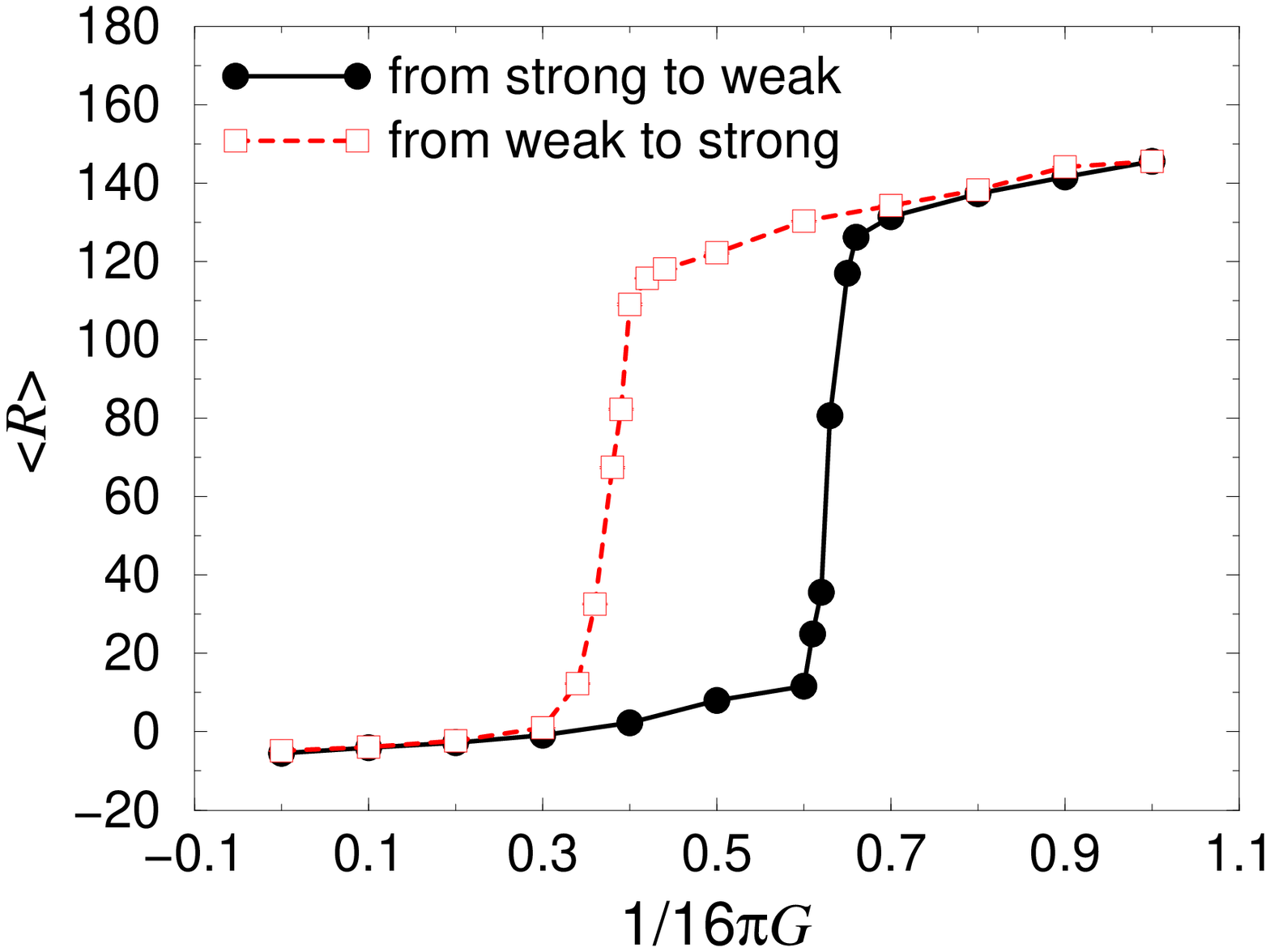}
\end{center}
\vspace*{-1.0cm}
\caption{{\sl
Hysteresis for the curvature $\langle R \rangle$ under the 
scale-invariant measure $\prod_i dl_i/l_i$. The solid line means the process 
from the strong to the weak coupling phase and the dashed vice versa.}}
\label{fig:ConMeasureCal}
\vspace{2.5cm}
\begin{center}
\includegraphics[height=8cm]{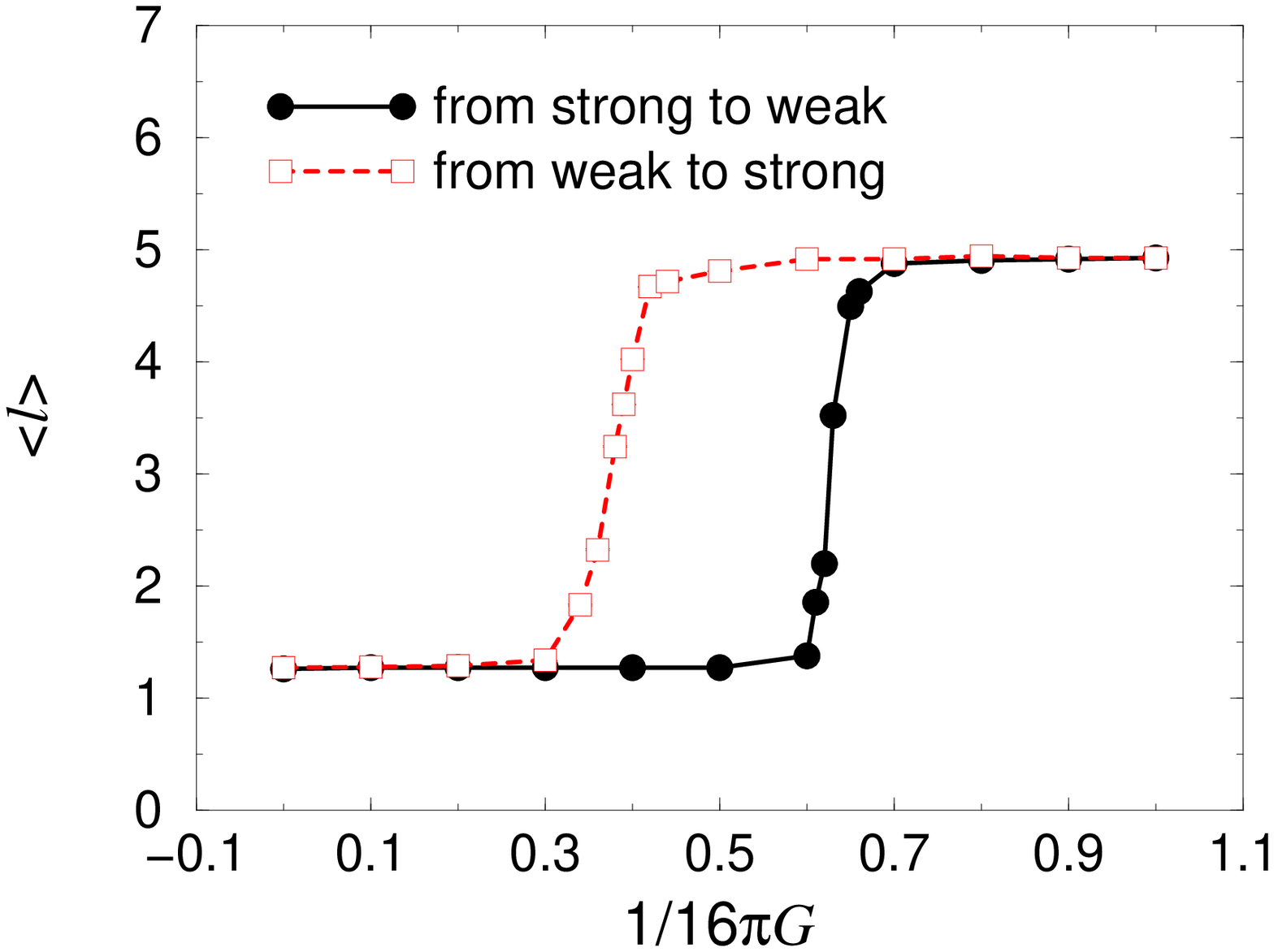}
\end{center}
\vspace*{-1.0cm}
\caption
{{\sl
Hysteresis for the average link-length $\langle l \rangle$ under the 
scale-invariant measure $\prod_i dl_i/l_i$. The solid line means the process 
from the strong to the weak coupling phase and the dashed vice versa.
}}
\label{fig:CM_length}
\vspace{0.0cm}
\end{figure}
%

  In this case, we can obtain in principle any observables numerically, 
unlike the case of the uniform measure. Indeed, we measured the 
scalar curvature per volume, $\langle R \rangle$, as shown in 
Fig.~\ref{fig:ConMeasureCal}. Evidently, this system has two phases, namely,
the strong and the weak coupling phases; the solid line represents the cooling 
process from the strong coupling phase to the weak one, and the dotted its 
inverse. One can clearly observe a large hysteresis in 
Fig.~\ref{fig:ConMeasureCal}, suggesting the first-order nature of the 
transition. It is consistent with results obtained in numerical studies of 
three-dimensional DT~\cite{AMBKEH}. 

  Next, we measured the average link-length $\langle l \rangle$ as shown in
Fig.~\ref{fig:CM_length}, where we observe another large hysteresis. In the 
strong-coupling phase, the strong gravitation prevents spacetime manifolds from
extending widely, resulting in the small value of the length 
$\langle l \rangle$. In the weak-coupling phase, conversely, spacetime 
manifolds tend to stretch as widely as possible, resulting in spiky 
configurations. Such spikes of spacetime manifolds often appear also in the 
weak-coupling phase of quantum RC~\cite{HW3}. In this sense, our 
data of Fig.~\ref{fig:CM_length} is consistent with those obtained in 
the numerical studies of quantum RC.

%
%
  Here we make a short remark on the statistics of our data.
To get each point in Figs.~\ref{fig:ConMeasureCal} and \ref{fig:CM_length},
we accumulate 200 configurations, each of which is obtained at an interval of 
800 sweeps. Statistical errors are also included in the figures, 
although they are too small to be visible.
%
%
\subsection{Fractal structure of 3D pure gravity}
\spc We further investigate other structures of spacetime manifolds under
the scale-invariant measure. Among them, a fractal structure based on a 
geodesic distance will be the most important, and its validity has been well 
established in two-dimensional DT~\cite{KKMW}. Here we first explain the 
fractal structure and related issues for our purpose.

\begin{figure}[h]
\vspace*{0.0cm}
\begin{center}
 \setlength{\unitlength}{1cm}
 \begin{picture}(12.5,10)(0,0)
 \put(1.5,0.5){\includegraphics[width=9.5cm]{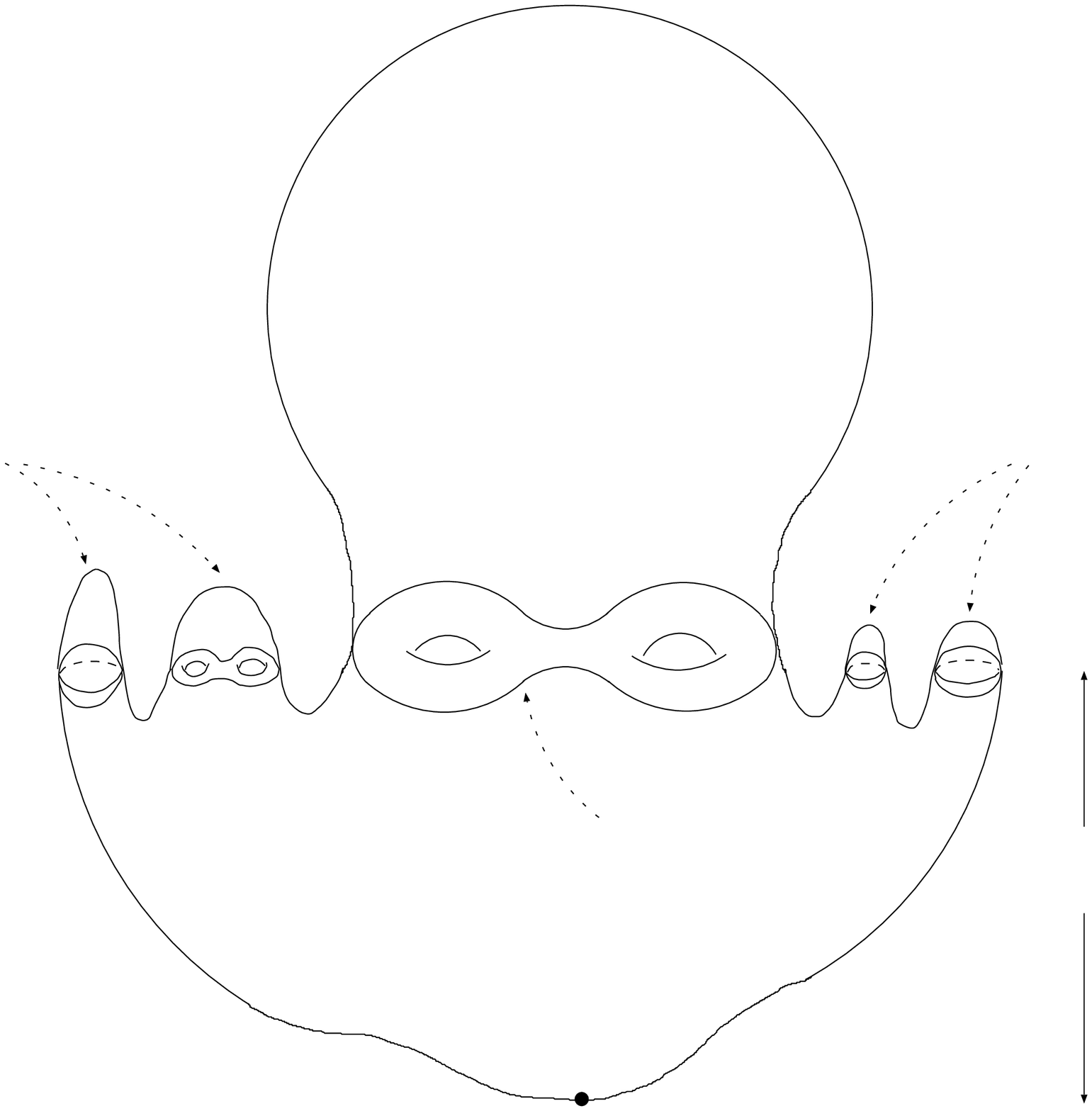}}
 \put(4.95,7.7){mother universe} \put(5.0,2.7){boundary of the mother}
 \put(0.3,6.3){baby universe} \put(9.6,6.3){baby universe}
 \put(10.8,2.45){$D$} \put(6.4,0.1){$P$}
 \end{picture}
\end{center}
\vspace*{-0.8cm}
\caption{{\sl
Schematic picture of a three-dimensional fractal manifold that appears in the 
strong-coupling phase of DT. On the `time slice' at a geodesic distance $D$
one can find many closed surfaces as the boundary. The fractal structure is
revealed as the scaling property of the distribution of the closed surfaces 
at various geodesic distances.
}}
\label{fig:frac_mfd_3d}
\vspace*{0.5cm}
\end{figure}

  In the case of 3D gravity, the fractal structure is characterized by the 
{\em surface area distribution (SAD) function} $\rho(S, D)$~\cite{HTY} 
and the {\em fractal dimension} $d_f$; their definitions will be given right 
now and closely related to each other.
The SAD function is defined as follows~\cite{HTY}: (i) Let us consider a 
connected graph $\tilde{T}$ dual to a three-dimensional closed simplicial  
manifold $T$; (ii) The {\it geodesic distance\/} between two points 
in $\tilde{T}$ is defined as the minimum number of steps between the two 
points; (iii) We pick up a point $P$ in the graph $\tilde{T}$ (namely, a 
3-simplex in $T$) and find all points that have a geodesic distance $D$ from 
the starting point $P$; (iv) The boundary manifold appearing in slicing $T$
at the geodesic distance $D$ consists of closed surfaces of various topologies,
as shown in Fig.~\ref{fig:frac_mfd_3d}; 
(v) Then, the SAD function $\rho(S,\ts D)$ is defined as 
\begin{eqnarray}
\rho(S,\ts D) \equiv \mbox{the number of closed surfaces of area $S$ at 
the geodesic distance $D$}.
\label{eq:SADfunc}
\end{eqnarray}

  The fractal structure is encoded in the scaling behavior of the 
SAD function $\rho(S, D)$ with a scaling variable $x = S / D^{\alpha}$, where 
the exponent $\alpha$ has different values in different phases. $\rho(S, D) 
\times D^{\alpha}$ is expected to be a function only of $x$ in the scaling 
regions, and, therefore, it suggests a scaling law $S \sim D^{\alpha}$. 
As a result, the fractal dimension, $d_f$, of spacetime is determined as
\begin{eqnarray}
d_f = \alpha + 1 \es.
\label{eq:frctlDim}
\vspace*{-0.1cm}
\end{eqnarray}
Intuitively, such a scaling means a self-similarity of `time slices' that arise
in cutting spacetime manifolds at different geodesic distances\footnote{
Actually, one can think of the geodesic distance $D$ as a time coordinate, 
corresponding to the so-called temporal gauge~\cite{FIKN} in 2D quantum
gravity.}.

  Incidentally, the scaling behavior of $\rho(S,\ts D)$ was observed 
numerically in three-dimensional DT~\cite{HTY}, while its validity in studying 
quantum spacetime was first established in two-dimensional DT~\cite{KKMW}. 
Moreover, a similar scaling behavior was reported also in four-dimensional 
DT~\cite{EHITY}.
%
%

  According to the values of the gravitational constant $G$, we can classify 
the fractal structure into three regions: the strong-coupling, the critical, 
and the weak-coupling regions. An example of the fractal manifold with the 
scaling behavior is schematically depicted on Fig.~\ref{fig:frac_mfd_3d}, which
occurs typically in the strong-coupling phase of three-dimensional DT. A closed
surface of fairly large area appears only once at each `time slice' $D$, which
is the boundary of the so-called {\em mother universe}. Concurrently, there are
many surfaces of small areas at each `time slice' and they are boundaries of 
the {\em baby universes}. The boundaries of the mother and the baby universes 
consist of closed surfaces of various topologies.
\vspace*{0.3cm}

  Having reviewed the fractal structure of lattice gravity, we will devote our 
attention to the investigation of the structure in our hybrid model.
%
\begin{figure}
\begin{center}
\includegraphics[width=10.5cm]{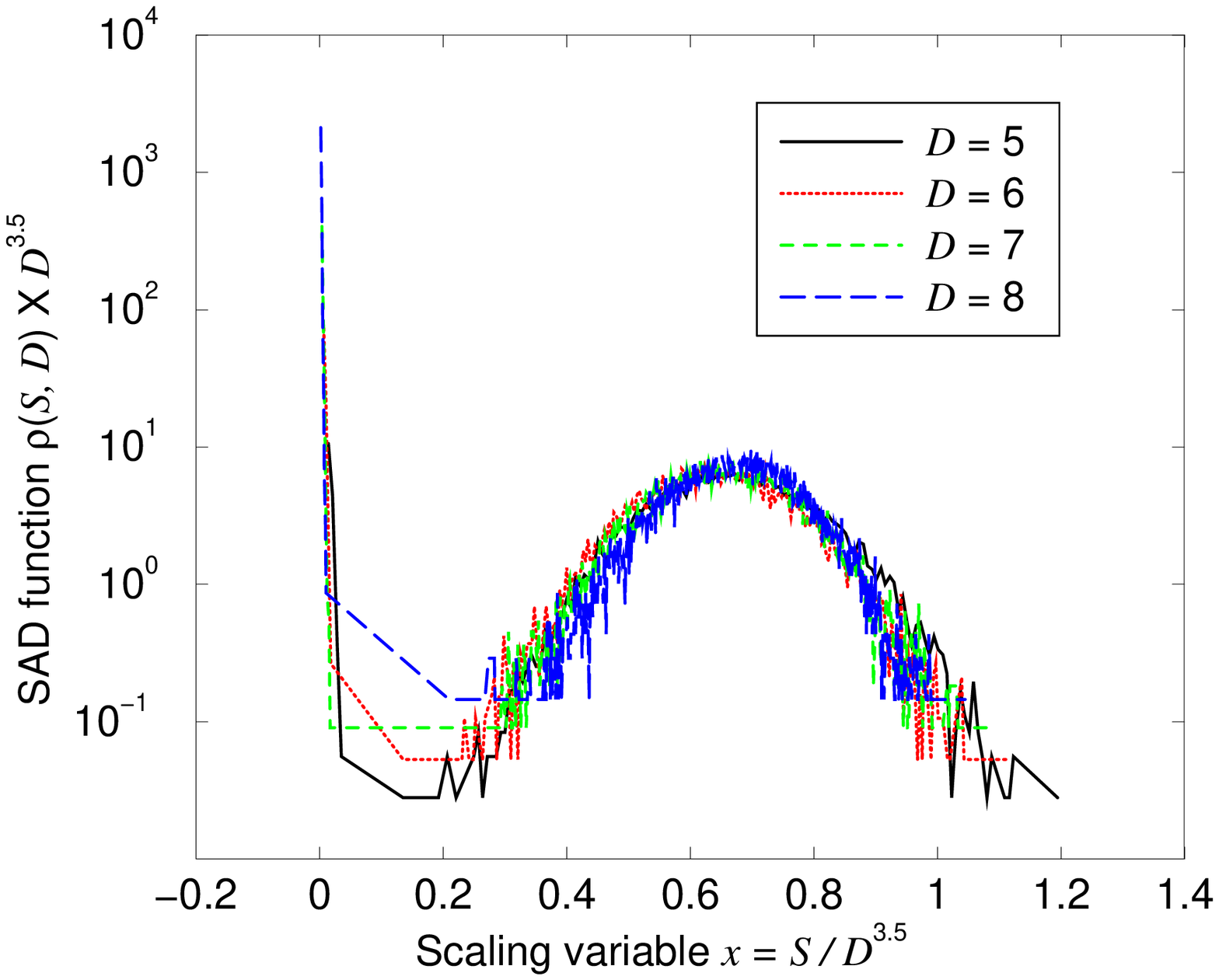}
\end{center}
\vspace*{-0.8cm}
\caption
{{\sl
SAD function $\rho(S,\ts D)$ in the strong-coupling limit $G = \infty$ 
of 3D pure gravity based on DRC. The number of 3-simplices are almost fixed at 
about $N_3 = 5000$ and each link-length $l_i$ is integrated over a range 
$1 < l_i < 10$. Different curves correspond to different geodesic distances 
$D = 5,6,7$ and 8. The scale-invariant measure $\prod_i d l_i /l_i$ is used.
}}
\label{fig:drc_ad_5k_cm_00_D5_8}
\vspace{1.5cm}
\begin{center}
\includegraphics[width=10.5cm]{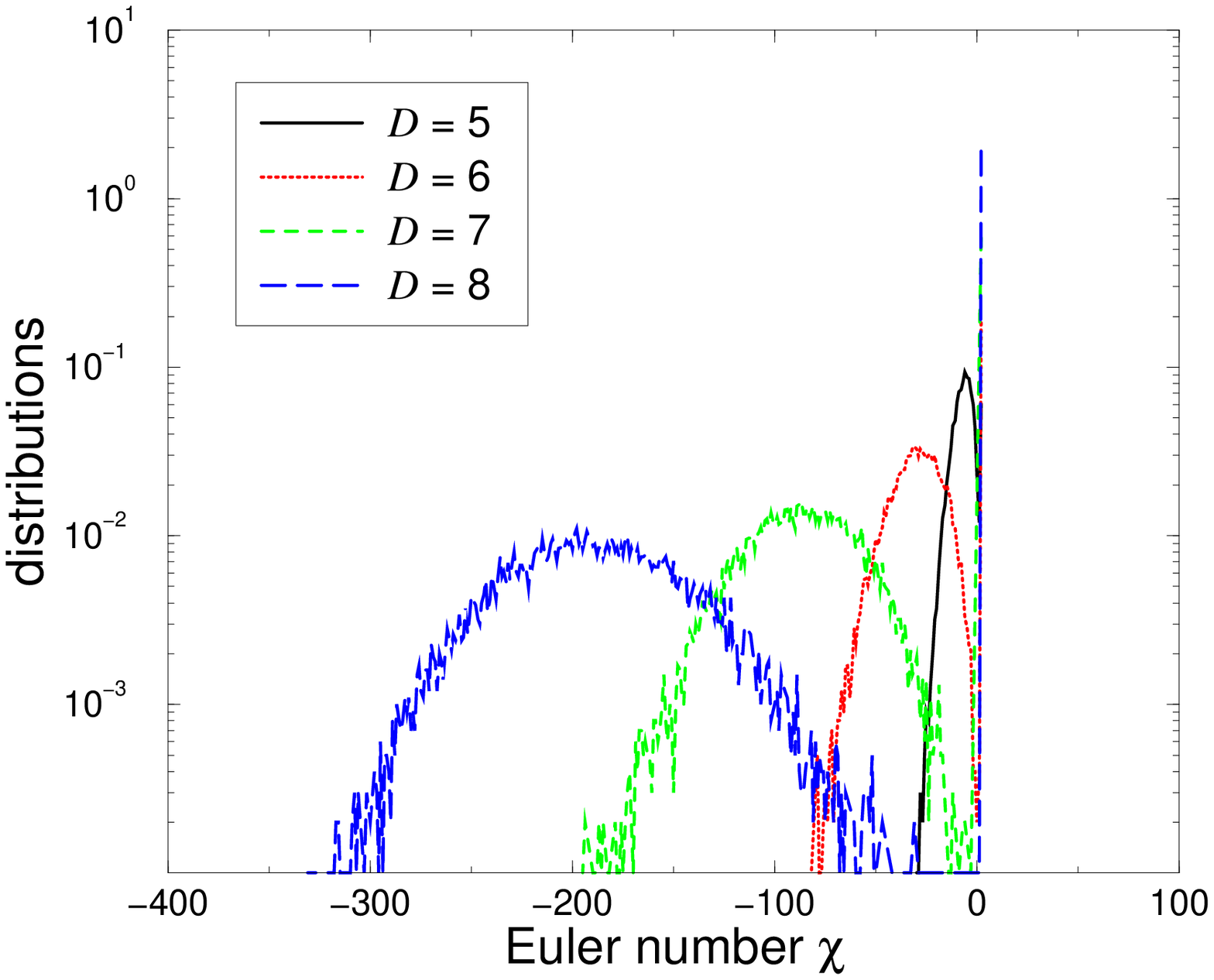}
\end{center}
\vspace*{-0.8cm}
\caption
{{\sl
Distributions of the Euler number $\chi$ in the strong-coupling limit 
$G = \infty$ of 3D pure gravity based on DRC. Different curves correspond to 
different geodesic distances $D = 5,6,7$ and 8. The scale-invariant measure 
$\prod_i d l_i /l_i$ is used.
}}
\label{fig:drc_ed_5k_cm_00_w6_rdc}
\vspace{-0.0cm}
\end{figure}
%
%
\subsubsection{Fractal structure in the strong-coupling limit $G = \infty$}
\spc First, we measure the SAD function $\rho(S,\ts D)$ in the strong-coupling 
limit $G = \infty$ in DRC. Numerical data obtained is shown in 
Fig.~\ref{fig:drc_ad_5k_cm_00_D5_8}, where the vertical axis means the
SAD scaling function of the form $\rho(S, D) \times D^{3.5}$ and
the horizontal a scaling variable $x = S / D^{3.5}$.
The different curves show several data measured at different geodesic 
distances $D = 5,6,7$ and 8. The boundary surface of the mother universe, which
has the largest area among all boundaries, has a good scaling property 
with the scaling parameter $x = S / D^{3.5}$. Accordingly, the fractal 
dimension is obtained as $d_f = 3.5 + 1 = 4.5$. One can clearly observe in 
Fig.~\ref{fig:drc_ad_5k_cm_00_D5_8} that the distribution of the mother 
universe has the Gaussian form. On the contrary, the distributions of the 
boundaries of the baby universes, whose areas are fairly smaller than that of 
the mother, do not exhibit a scaling behavior. In other words, it is impossible
that the baby part scales in the same way as the mother does; such phenomenon
of the existence of two scaling variables in a simplicial manifold typically 
appear also in three-dimensional DT~\cite{HTY}.

  Next, we measure the Euler number of the boundary surfaces. On each closed 
surface, the Euler number $\chi$ is defined in the usual way:
\begin{eqnarray}
\chi \equiv n_0 - n_1 + n_2 \es,
\label{eq:EulerNumber_ed_5k_cm_00_w6_rdc}
\end{eqnarray}
where $n_k$ mean the numbers of $k$-simplices $(k = 0,1,2)$ on each boundary
surface. Our numerical data of $\chi$ are shown in 
Fig.~\ref{fig:drc_ed_5k_cm_00_w6_rdc}; obviously, surfaces with large negative 
$\chi$, exhibiting complicated topologies, are identified with the boundary of 
the mother universe. The distributions of the mother are in the shape of 
mountain (Gaussian) and, in contrast, those of the baby universes have a sharp 
peak at $\chi = 2$. This result is consistent with that obtained in the 
strong-coupling phase of three-dimensional DT~\cite{HTY}.

  How can we draw a physical picture of this region? A plausible answer is a 
`confinement of spacetime', into which all the volume of quantum spacetime is 
completely confined\footnote{If one can obtain the `confinement' picture also 
in four dimensions, it would be regarded as the confinement of gravitons. 
However, we are now in three dimensions, where no gravitons exist.}.
Actually, the large fractal dimension, whose value is $d_f \sim 3.5+1$
in this limit, will prevent spacetime from extending, although the volume 
is not small. In addition, the large negative values of $\chi$ frequently 
appear in the distributions of the mother; this complexity of the topology
reminds us of the spacetime foam.
%
%
%
\subsubsection{Fractal structure in the critical region $G \sim G_c$}
\begin{figure}
\vspace{0.0cm}
\begin{center}
\includegraphics[width=10.5cm]{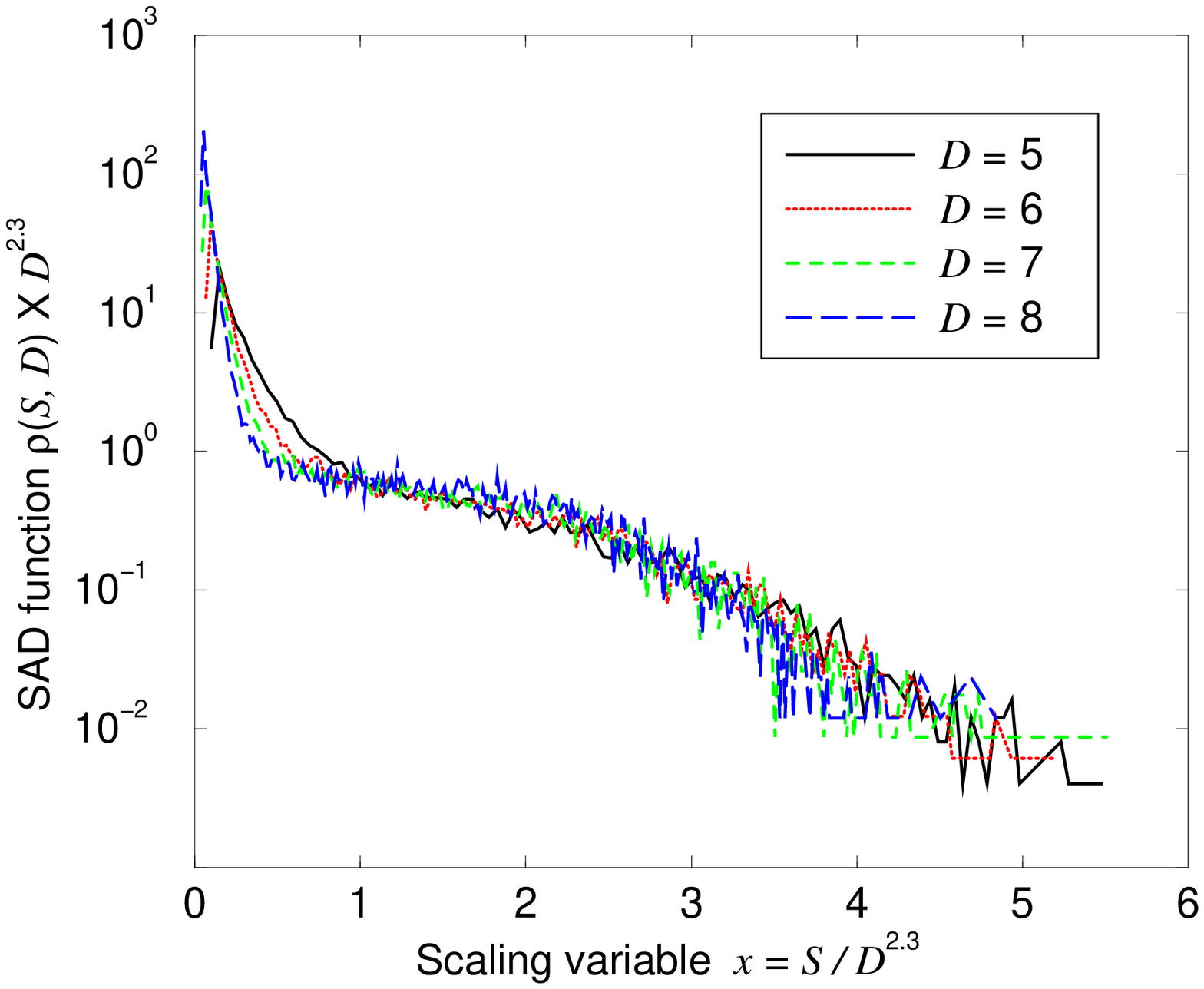}
\end{center}
\vspace*{-0.8cm}
\caption
{{\sl
SAD function $\rho(S,\ts D)$ at $1/16\pi G = 0.6$ in the critical region 
of 3D pure gravity based on DRC. The number of 3-simplices are almost fixed at 
about $N_3 = 5000$ and each link-length $l_i$ is integrated over a range 
$1 < l_i < 10$. Different curves correspond to different geodesic distances 
$D = 5,6,7$ and 8. The scale-invariant measure $\prod_i d l_i /l_i$ is used.
}}
\label{fig:drc_ad_5k_cm_060_D5_9}
\vspace{1.5cm}
\begin{center}
\includegraphics[width=10.5cm]{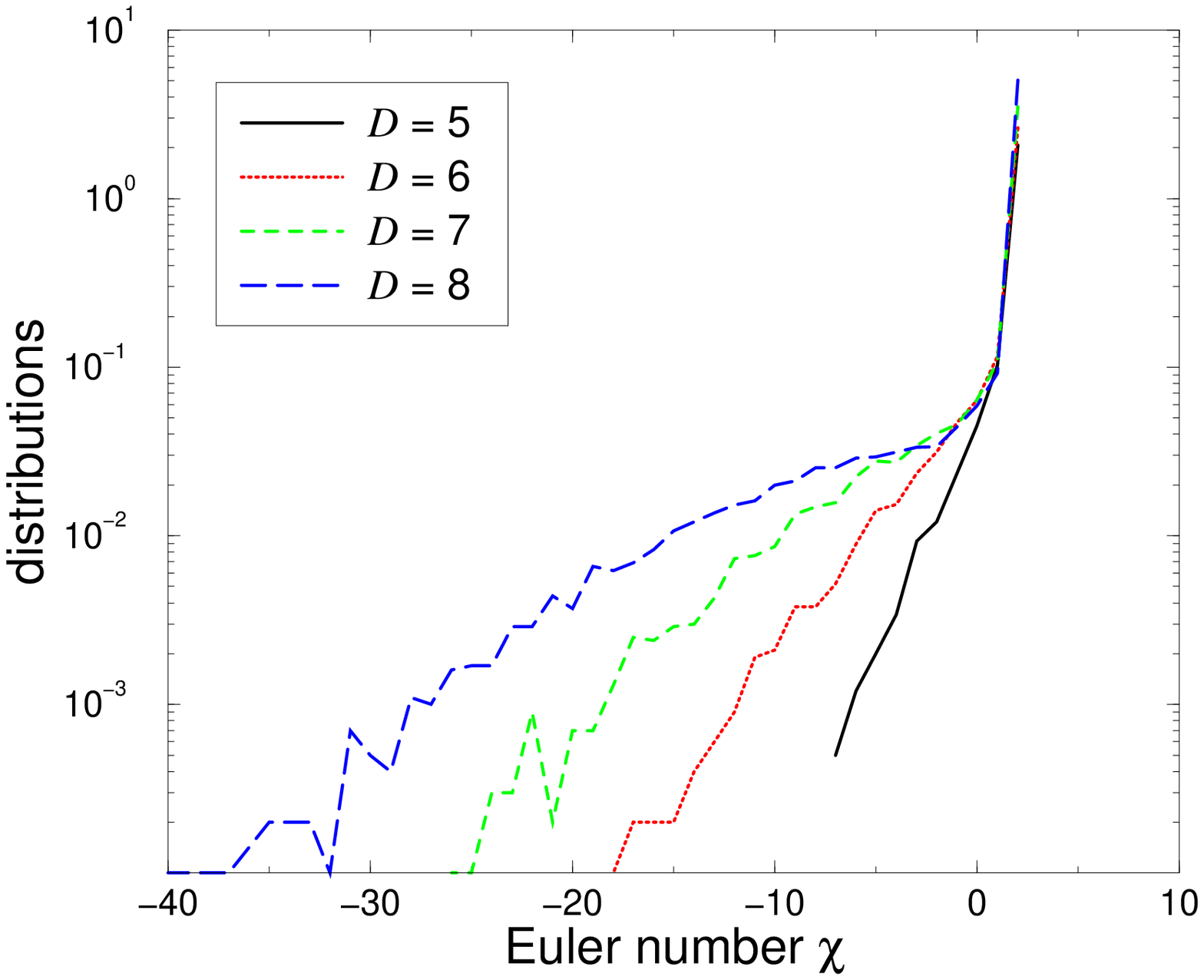}
\end{center}
\vspace*{-0.8cm}
\caption
{{\sl
Distributions of the Euler number $\chi$ at $1/4\pi G = 0.6$ in the critical 
region of 3D pure gravity based on DRC. Different curves corresponds to 
different geodesic distances $D = 5,6,7$ and 8. The scale-invariant measure 
$\prod_i d l_i /l_i$ is used.
}}
\label{fig:drc_ed_5k_cm_060_s9}
\vspace{-0.0cm}
\end{figure}

\spc Second, we measure the SAD function and the Euler number distribution
at $1/16\pi G = 0.6$ in the critical region. Our numerical data are shown in 
Fig.~\ref{fig:drc_ad_5k_cm_060_D5_9} and Fig.~\ref{fig:drc_ed_5k_cm_060_s9}.
In Fig.~\ref{fig:drc_ad_5k_cm_060_D5_9}, the vertical axis is
the scaling function of the form $\rho(S, D) \times D^{2.3}$ and the horizontal
the scaling variable $x = S / D^{2.3}$; the exponent 2.3 is vary small 
compared with the value obtained in the strong-coupling limit. Thus, we obtain 
a smaller fractal dimension $d_f = 2.3 + 1 = 3.3$ in this region. In general, 
the smaller the fractal dimension becomes, the more widely the spacetime 
manifold extends.
%

  In this region, one cannot separate the contribution of the mother from that 
of the baby universes, because the scaling function $\rho(S,\ts D)$ 
distributes smoothly from small $x$ to large one. We have learned in 
two-dimensional DT that the distribution of the mother boundary is
universal and, hence, it has the close connection with the continuum 
limit~\cite{KKMW}. On the other hand, the distribution of the baby
boundaries is non-universal; interestingly, the result shown in 
Fig.~\ref{fig:drc_ad_5k_cm_060_D5_9} is very similar to that obtained
in two-dimensional DT~\cite{KKMW}, although the first-order nature of
the transition indicates no continuum limit in our case.

  Next, we measure the $\chi$ distributions of closed surfaces that appear in 
cutting the simplicial lattice at each geodesic distance. Our data is shown in 
Fig.~\ref{fig:drc_ed_5k_cm_060_s9}, where the Euler number $\chi$ is measured 
at several distances. One can see that the $\chi$ distributions are fairly 
smoother than those of Fig.~\ref{fig:drc_ed_5k_cm_00_w6_rdc}. As a
result, we get a milder picture of spacetime in this region and
the shape of the $\chi$ distribution is very similar to that obtained in 
three-dimensional DT~\cite{HTY}. 
%
%
%
\subsubsection{Fractal structure in the weak-coupling phase $G < G_c$}
\spc Lastly, in the weak-coupling phase, we attempted to measure the same 
quantity $\rho(S,\ts D)$ in the same manner as in the strong-coupling
and the critical regions. However, we cannot find any contributions
from the mother universe in this phase and, what is worse, lattice 
configurations seem to be almost frozen, which have a lower fractal
dimension; this phenomenon strongly suggests the spiky nature of
spacetime in this phase, and such a spike is far from the usual notion
of physical spacetime. As is well known with numerical studies of DT,
this phase corresponds to an elongated branched-polymer with a small
fractal dimension (for example, $d_f = 2$)~\cite{HTY}.
Similarly, such a spike has been observed also in Monte-Carlo studies of
quantum RC~\cite{HW3}.
\vspace*{7mm}
\section{Conclusions and discussion}\label{sect:sumryDscsn}
\spc We have proposed dynamical Regge calculus~(DRC) as a hybrid model of 
simplicial quantum gravity. This model is intended to make physical degrees of 
freedom larger than those of quantum Regge calculus~(RC) and  dynamical 
triangulations~(DT). Furthermore, the extended model of DRC gives the 
possibility of describing the topology-changing processes of Euclidean 
spacetime in a dynamical way, although there are some difficulties in 
simulating the processes numerically.

  Algorithmically, the path integral for DRC (\ref{eq:Z_DRC}) can be performed 
through the hybrid $(p,\ts q)$ moves, which are an extension of the ergodic 
$(p,\ts q)$ moves of DT. In particular, the lattice diffeomorphisms are 
generated by the invariance $(p,\ts q)$ moves.
It is also an interesting problem that other constructive approaches to quantum
gravity reproduce the same structure of the lattice diffeomorphisms as that of 
DRC, if one would believe the universality of field theory also in 
quantum gravity.

%
%
  As an application of the lattice diffeomorphism, we tried a 
lattice-theoretic derivation of the black hole entropy. We identified 
the total number of quantum fluctuations around the event horizon with the 
black hole entropy; such hybrid moves that keep invariant the horizontal 
geometry play the important role. In order to avoid the divergence of the 
coefficient of the entropy, we required the lattice cutoff $l_{\rm min}$ to 
remain a finite value of order $O(\ell_{\rm P})$. The introduction of the 
minimal length seems to be consistent with a simple interpretation of the 
space-time uncertainty principle of string theory.

  Moreover, we have carried out the numerical simulations of 3D pure gravity 
using the two kinds of the integration measures. 
In case of the uniform measure $\prod_i dl_i$ we observed the divergence 
behavior of the lattice size $N_3$ even though we chose the large values of
the cosmological constant. This phenomenon indicates the exponential 
unboundedness for the entropy owing to the lattice diffeomorphisms. 
In this case DRC is not well-defined statistical-mechanically.

  In case of the scale-invariant measure $\prod_i dl_i/l_i$, however, no
pathological behavior occurred in our data. Indeed, we calculated the average 
curvature $\langle R \rangle$ and the average link-length $\langle l \rangle$; 
two pieces of large hysteresis were observed, indicating the existence of the 
first-order transition between the two phases. In the strong-coupling limit, 
spacetime manifolds are crumpled to a singular configuration with a large 
fractal dimension. 
Physically, such a crumpled space might suggest the existence of `confinement',
into which spacetime itself is confined.
In addition, various topologies appear on time slices of `confined spacetime'. 
%
  In the critical region, we have obtained the smooth SAD function,
resulting in a milder picture of spacetime than in the strong-coupling limit. 
In this case spacetime looks a fractal manifold of lower fractal dimension.
  On the other hand, in the weak coupling phase simplicial configurations 
become spiky, and this phenomenon is essentially the same as
that has been observed in the weak-coupling phase of quantum RC.
Taking into account these theoretical and numerical studies, we conclude 
that DRC can reproduce the numerical results consistent with those of both DT 
and quantum RC in each region.

  Our hybrid model of lattice quantum gravity offers a practical way of 
studying quantum black hole physics and the topology-change of Euclidean 
spacetime on the lattice. Incidentally, the close relation between 3D quantum 
gravity and quantization of the membrane theory~\cite{membrane} is an 
interesting theme in connection with string theory, though such an attempt
generally gives rise to an instability problem~\cite{WHN}.
Our chief concern is to investigate numerically the physics of strong-coupling 
quantum gravity within the framework of (extended) DRC. 
\vspace*{1cm}

\noindent{\large {\bf Acknowledgments}}

  I am grateful to K.~Hamada for his detailed comments and suggestions.
I would like to thank G.~Kang for his useful comments on quantum black holes.
I also wish to thank T.~Yukawa and S.~Horata for many useful discussions.
Finally, I want to acknowledge valuable discussions with many members of 
Theory Division of KEK.

%
%
%
%

\end{document}